\documentclass[aps,prl,preprintnumbers,showpacs,nofootinbib]{revtex4}
\usepackage[utf8]{inputenc}
\usepackage[english]{babel}
\usepackage{amsmath}
\usepackage{amsfonts}
\usepackage{amssymb}
\usepackage{color}
\usepackage{slashed}
\usepackage{enumerate}
\usepackage{graphicx}
\usepackage{bm}
\usepackage{braket}
\usepackage{epstopdf}
\usepackage[usenames,dvipsnames,svgnames]{xcolor}
\usepackage[colorlinks=true,
            linkcolor=red,
            urlcolor=gray,
            citecolor=blue]{hyperref}

\providecommand{\NEG}{\slashed}
\begin{document}

\title{%
\textcolor{blue}{\textbf{Phenomenology of the new physics coming from
2HDMs to the neutrino magnetic dipole moment.}}}
\author{Carlos G. Tarazona$^{(a,b)}$}
\author{Rodolfo A. Diaz$^{(a)}$}
\author{John Morales$^{(a)}$}
\author{Andr\'es Castillo$^{(a)}$}
\affiliation{$^{(a)}$ Universidad Nacional de Colombia, Sede Bogot\'a, Facultad de
Ciencias, Departamento de F\'{\i}sica. Ciudad Universitaria 111321,
Bogot\'a, Colombia}
\affiliation{$^{(b)}$Departamento de Ciencias B\'asicas. Universidad Manuela Beltr\'an.
Bogot\'a, Colombia}
\date{\today }

\begin{abstract}
In several frameworks for leptons-sectors of two Higgs doublet models, we
calculate the magnetic dipole moment for the different flavor types of
neutrino. Computations are carried out by assuming a normal hierarchy for
neutrino masses, and analyzing the process $\nu \rightarrow \nu \gamma $
with a charged Higgs boson into the loop. The analysis was performed by
sweeping the charged Higgs mass and taking into account the experimental
constraints for relevant parameters in Two Higgs Doublet Models with and
without flavor changing neutral currents; obtaining magnetic dipole moments
close to the experimental thresholds for tau neutrinos in type II and
Lepton-specific cases. In the neutrino-specific scenario, the contribution
of new physics could be sizeable to the current measurement for flavor
magnetic dipole moment. This fact leads to excluding possible zones in the
parameter space of charged Higgs mass and vacuum expectation value of the
second doublet.
\end{abstract}

\maketitle

%%%%%%%%%%%%%%%%%%%%%%%%%%%%%%%%%%%%%%%%%%%%%%%%%%%%%%%%
%%%%%%%%%%%%%%%%%%%%%%%%%%%%%%%%%%%%%%%%%%%%%%%%%%%%%%%%
%%%%%%%%%%%%%%%%%%%%%%%%%%%%%%%%%%%%%%%%%%%%%%%%%%%%%%%%
%%%%%%%%%%%%%%%%%%%%%%%%%%%%%%%%%%%%%%%%%%%%%%%%%%%%%%%%


\vspace{0.1cm}

\section{Introduction}

All elementary charged fermions in the Standard Model (SM) are Dirac
fermions. Nevertheless, the nature of the neutrino is not yet definitely
settled and depending on the model the neutrino can be either a Majorana or
Dirac fermion. By discriminating the electromagnetic behavior of the
neutrino, we could have an alternative resource (aside the neutrinoless
double beta decay) to determine the nature of such a particle. Since
neutrinos do not carry electric charge, they can participate in
electromagnetic interactions by coupling with photons only via quantum
corrections. Likewise, as it happens for other particles, electromagnetic
properties of neutrinos can be described employing electromagnetic form
factors (EFFs). For example, by the realization of their multipole moments,
neutrinos can be sensitive to intense electromagnetic fields. Such intense
fields can exist in nature indeed; it has been suggested that there could be
sources of magnetic fields of order $\left( 10^{13}-10^{18}\right) $ G, as
it could be the case during a supernova explosion or in the vicinity of
particular groups of neutron stars known as magnetars \cite{Mereghetti}.

Several experiments have measured phenomenological constraints over
neutrinos EFFs, particularly their Magnetic Dipole Moment (MDM), which in
the most stringent bound the respective value should be less to $10^{-11}$ $%
\mu _{B}$ \cite{Mohapatra, Giunti, Kouzakov}. The first prototypical model
to take into account these experimental thresholds has been the SM with
right-handed neutrinos (transforming as singlets of SM gauge group).
However, and from effective operators computations, MDMs in this minimal
extended SM give contributions with considerable several orders of magnitude
below of this experimental threshold; under the small values of neutrino
masses \cite{Alberto}. A way to enhance these contributions can be
introduced by new physics effects \cite{Balantekin}, such as those described
by a well motivated extended Higgs sector.

Present limits on the scalar sector in the SM still allow the possibility of
an extended Higgs sector, if the correspondent effects are weak enough in
the current decay channels for the detection of the SM like scalar in the
mass region around of $125$ GeV \cite{G. Aad}. We shall study one of the
simplest extensions of the scalar sector of the SM, the so-called Two Higgs
Doublet Model (2HDM) in which we add to the symmetry breaking sector of
electroweak gauge group two Higgs doublets with the same quantum numbers of
isospin and hypercharge. There are many motivations for this model. One of
them is the fact that the SM with just one Higgs doublet is unable to
generate a baryon asymmetry of the universe of sufficient size as well as
plausible dark matter candidates, or to explain the mass hierarchy in the
third generation of quarks. Two Higgs Doublet Models are possible scenarios
to solve these problems, due to the flexibility of their scalar mass
spectrum and the existence of additional sources of $CP$ violation \cite%
{Diaz}. Also, in the Minimal Supersymmetric Standard Model (MSSM), a second
doublet should be added to cancel chiral anomalies \cite{Sher}; since
scalars are represented by chiral multiplets together with spin $1/2$
fields. Moreover, in MSSM, a minimal Higgs sector is unable to give the mass
to the up type quarks and down type quarks simultaneously, because doublets
of different chiralities cannot be coupled together in the Lagrangian. Thus,
a second Higgs doublet must be introduced to endow all quarks with masses.

Furthermore, the origin of neutrino masses is still an open question in High
energy physics. Explaining the smallness of neutrino masses in the SM
demands to consider an effective operator involving a possible High energy
scale for new physics related to perhaps a most fundamental theory. Indeed,
2HDMs could be invoked to describe the neutrino-hierarchy problem by the
introduction of a Vacuum Expectation Value (VEV) for one of the doublets in
the scale of neutrino masses. This approach can be achieved by the
realization of 2HDM-neutrino specific, where fundamentals of the model are
incorporated when the new doublet couples only to the neutrino sector \cite%
{Gabriel,logan1,logan2}.

When 2HDM is built up, the general form of Yukawa couplings among fermions
and scalars compatible with gauge invariance have rare processes called
Flavor Changing Neutral Currents (FCNCs). Experiments such as $\bar{K}%
^{0}-K^{0}$ mixing highly constrain flavor violation currents \cite{Gupta,
Gupta5}. The discrete symmetry $Z_{2}$ (e.g. $\Phi _{1}\leftrightarrow \Phi
_{1}$ and $\Phi _{2}\leftrightarrow -\Phi _{2}$) as an intrinsic parity is
usually implemented in the 2HDM because it forbids mixing between the two
doublets, which is the primary source of FCNCs. Moreover, this symmetry
ensures a CP-conserving frame in the scalar sector when $Z_{2}$ symmetry
extended to the Higgs potential. In the last point, the $Z_{2}$
transformation unfolds the 2HDM since its presence (absence) leads to
different forms of the Yukawa couplings between fermions and Higgs bosons
with (without) flavor natural conservation. Besides to this discrete
symmetry, a continuous global symmetry like $U(1)$ could be incorporated to
achieve this FCNC suppression. These symmetry implementations respect the
so-called Weinberg-Glashow theorem: all fermions should be coupled at most
to one doublet to avoid FCNCs naturally. In our description of neutrino
MDMs, we consider three models with suppression of FCNCs at tree level (type
I, II, flipped, lepton and neutrino-specific 2HDMs) and one model with that
kind of couplings, i.e., the type III-2HDM.

As it was pointed out above, couplings of neutrinos with photons occur via
loop diagrams. In the Standard Model (SM), the loop corrections have the
form of vertex diagrams and vacuum polarization diagrams. When the SSB
sector includes a second doublet, further corrections appear by replacing
the vector bosons $W^{\pm }$ by charged Higgs bosons $H^{\pm }$. Our goal is
to characterize the corrections to the EFF's coming from the new physics and
particularly in the region of parameters in which such factors become near
to the threshold of experimental detection.

The structure of this paper is as follows. We discuss the general form of
the EFF's for neutrinos in section \ref{form factors}. Subsequently, in
section \ref{2HDM} we review the two Higgs doublet Model (2HDM) and its
possible realizations to incorporate the neutrino masses. In section \ref%
{nspecific} we include the neutrino-specific model with viable mass terms
for neutrinos. We dedicate section \ref{loops} to study the behavior of the
form factors for these different 2HDMs. In section \ref{results} we perform
the correlations between the masses and the possibility of finding
observable effects coming from EFF's of the neutrino in the different
frameworks of 2HDMs with natural flavor conservation and under the presence
of FCNCs. Finally, section \ref{sec:conclusions} highlights our conclusions
and remarks, as well as perspectives to implement these constraints in other
models with new physics beyond SM.

\section{The electromagnetic form factors (EFF's)\label{form factors}}

\begin{figure}[htp]
%[tph]
\centering
\includegraphics[scale=1.1]{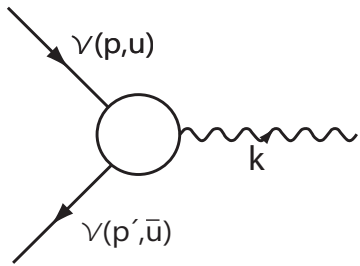} \vspace{0.3cm}
\caption{\textit{Effective coupling representation of two neutrinos with a
photon.}}
\label{vertex}
\end{figure}

In this section, we review the general structure of the electromagnetic
interactions of Dirac and Majorana neutrinos in the one-photon
approximation. To find all the EFF's, we use the general expression for the
electromagnetic current

\begin{equation}
\left\langle u\left( p,\lambda \right) |J_{\mu }^{EM}\left( x\right)
|u\left( p^{\prime },\lambda ^{\prime }\right) \right\rangle =\overline{u}%
\left( p,\lambda \right) \Lambda _{\mu }\left( l,q\right) u\left( p^{\prime
},\lambda ^{\prime }\right) ,
\end{equation}

where $q_{\mu }=p_{\mu }^{\prime }-p_{\mu }$, $l_{\mu }=p_{\mu }^{\prime
}+p_{\mu }$ are the four-momenta shown in Fig. \ref{vertex}, and\ $u\left(
p,\lambda \right) $, $u\left( p^{\prime },\lambda ^{\prime }\right) $ are
the initial and final fermion states respectively.

Further, $\Lambda _{\mu }$ are matrices of couplings acting on the spinors.
The matrices $\Lambda _{\mu }$ have some interesting properties \cite%
{Nova,broggini,studenikin}

\begin{itemize}
\item The first condition is that the arrangement $\Lambda _{\mu }$ must be
a cuadrivector, i.e. must be Lorentz covariant.

\item The second condition is hermiticity of the associated current, i.e., $%
J_{\mu }^{^{\dag }EM}=J_{\mu }^{EM}$ which implies%
\begin{equation}
\Lambda _{\mu }\left( l,q\right) =\gamma ^{0}\Lambda _{\mu }^{\dag }\left(
l,-q\right) \gamma ^{0}.
\end{equation}

\item The current conservation or gauge invariance $\partial ^{\mu }J_{\mu
}^{EM}=0$ gives%
\begin{equation}
q^{\mu }\overline{u}\left( p^{\prime },\lambda ^{\prime }\right) \Lambda
_{\mu }\left( l,q\right) u\left( p,\lambda \right) =0.
\end{equation}
\end{itemize}

Finally, the most general expression for $\Lambda _{\mu }\left( l,q\right) $
reads%
\begin{equation}
\Lambda _{\mu }\left( q\right) =F_{Q}\left( q^{2}\right) \gamma _{\mu }+ 
\left[ F_{M}\left( q^{2}\right) i+F_{E}\left( q^{2}\right) \gamma _{5}\right]
\sigma _{\mu \nu }q^{\nu }+F_{A}\left( q^{2}\right) \left( q^{2}\gamma
_{\mu}-q_{\mu }\neg{q}\right) \gamma _{5},  \label{Fullvertex}
\end{equation}%
where $F_{Q},~F_{M},~F_{E}$ and $F_{A}$ represent the electric charge,
magnetic dipole moment, electric dipole moment and anapole moment
respectively.

The EFF's show us how the particles couple with the photon at the tree level
or in loop corrections. At the tree level, we got the electric charge and
one part of the contribution coming from the magnetic dipole moment. Now, if
we consider the interaction with an external field $A_{ext}^{\mu }$ in the
form 
\begin{equation}
\mathcal{L}_{ext}=-eA_{ext}^{\mu }J_{\mu }^{EM},
\end{equation}%
the so-called anomalous magnetic moment arises. Even uncharged particles may
have a magnetic dipolar moment. However, for uncharged particles, all dipole
moments only appear in loop corrections. Just like the anomalous magnetic
moment, the electric dipole moment and the anapole moment can be non-zero
even for an uncharged particle \cite{lepton}.

In a similar way to what happens to other particles, neutrinos can be
described by EFF's with vertex functions. For neutrinos, the magnetic and
electric dipole moments are expected to be slight since they are likely
proportional to the neutrino masses. For instance, the leading contribution
to anomalous magnetic moment is \cite{lepton}%
\begin{equation}
a_{\nu _{i}}=-\frac{3G_{F}m_{\nu _{i}}}{4\sqrt{2}\pi ^{2}}m_{e}.
\end{equation}%
Consequently, the neutrino magnetic moment is%
\begin{eqnarray}
\vec{\mu}_{\nu _{i}} &=&\frac{e}{m_{e}}a_{\nu _{i}}\vec{s}_{\nu _{i}}  \notag
\\
&\Rightarrow &\Lambda _{SM}\equiv \mu _{\nu _{i}}=\frac{3G_{F}em_{\nu _{i}}}{%
4\sqrt{2}\pi ^{2}}\simeq 3.2\times 10^{-19}\left( \frac{m_{\nu _{i}}}{1\text{
eV}}\right) \mu _{B}.  \label{MDMSM}
\end{eqnarray}%
This relation was derived for the first time in \cite{Fujikawa}, where $\mu
_{B}$ is the Bohr's magneton. Other systematics, as the background field
method and computations based on neutrino self-energy, have determined
values in the same structure and the same order of magnitude for MDM of
neutrino \cite{Background,Kuznetsov,Erdas}. If neutrino couples to photons
via such moments, the neutrino electromagnetic properties can be used to
distinguish Majorana and Dirac neutrinos. For Dirac neutrinos, the most
important moment is $F_{M}$ because the other terms vanish in a $CP-$%
conserving scenario with a Hermitian $J_{\mu }^{EM}$, and are highly
suppressed owing to the soft violation of $CP$. On the other hand, for
Majorana neutrinos only $F_{A}$ is possible because the other terms vanish
due to the self-conjugate nature of Majorana neutrinos. Table \ref{tab:EMNC}
summarizes the contribution for the MDM of massive neutrinos (with effective
flavor masses). 
\begin{table}[tbph]
\begin{center}
\begin{equation*}
\begin{tabular}{|l|l|l|}
\hline
$l$ & Mass$\left( \frac{eV}{c^{2}}\right) $ & Magnetic dipole moment $%
\Lambda _{SM}$ \\ \hline
$\nu _{e}$ & $0.06089$ & $1.948\times 10^{-20}\mu _{B}$ \\ \hline
$\nu _{\mu }$ & $0.06754$ & $2.161\times 10^{-20}\mu _{B}$ \\ \hline
$\nu _{\tau }$ & $0.07147$ & $2.287\times 10^{-20}\mu _{B}$ \\ \hline
\end{tabular}%
\end{equation*}%
\end{center}
\par
\vspace{-0.4cm}
\caption{\textit{Magnetic dipole moments of neutrinos in the SM scenario.
Flavor effective masses are values compatible with central values for PMNS
matrix \protect\cite{Forero} and with cosmological bounds \protect\cite%
{Abazajian,PAde} and differences over masses for $\protect\nu_{1},\protect\nu%
_{2}$ and $\protect\nu_{3}$ eigenstates in normal ordering \protect\cite{PDG}%
. The effective values for masses, obtained in this way, satisfy the bounds
for the average masses of the flavor eigenstates $m_{\protect\nu_{e}},m_{%
\protect\nu _{\protect\mu }},m_{\protect\nu _{\protect\tau }}<2.5$ eV 
\protect\cite{Fukugita}, and $m_{\bar{\protect\nu}_{e}}<2.05$ eV 
\protect\cite{PDG}.}}
\label{tab:EMNC}
\end{table}

Constraints over neutrino MDMs are a result of scattering experiments (based
on mainly on the distortion of the recoil of charged leptons energy
spectrum) \cite{ValleNew}. In these experiments, the flavor neutrino
produced at some distance from the detector is a superposition of neutrino
mass eigenstates. Therefore, the MDM measured is an effective value which
takes into account neutrino mixing and the oscillations during the
propagation between source and detector \cite{broggini}. All computations
shall be referred to an effective magnetic moment of a flavor neutrino
without indication of a source-detector distance $L$. Indeed, it is
implicitly understood that this value is small, such that the effective
magnetic moment is independent of the neutrino energy and from the source
detector distance \cite{studenikin}. In such case, the effective MDM is $%
\mu_{\nu_{l}}^{2}\simeq \mu_{\bar{\nu}_{l}}^{2}\simeq \sum_{j=1}^{3}\vert
\sum_{i=1}^{3}U_{li}^{*}\left(\mu_{ji}-i\epsilon_{ji}\right)\vert^{2}$;
being $\mu_{ji}$ and $\epsilon_{ij}$ the neutrino magnetic dipole moment and
electric dipole moment respectively.

In a CP-conserving case, Majorana neutrinos have only transition elements ($%
F_{M}, F_{E}$ are antisymmetric implying Majorana neutrinos do not have
diagonal elements). Nonetheless, flavor neutrinos can have effective
magnetic moments even if neutrinos are of Majorana nature. In this case,
since massive Majorana neutrinos do not have diagonal magnetic and electric
dipole moments, the effective magnetic moments of flavor neutrinos receive
contributions only from the transition dipole moments.

Henceforth, we shall only be focused in Dirac neutrinos because considered
models in the next section introduce right-handed neutrinos in such a way
Lepton number is conserved. These scenarios based on different Yukawa
sectors of 2HDMs lead us to elucidate as effective MDMs for flavor Dirac
neutrinos arise regarding new physics parameters.

\section{The two Higgs doublet model with massive neutrinos\label{2HDM}}

Before introducing new physics effects in MDMs, we review several 2HDMs
differentiating flavor properties; which are relevant in the form of
interpreting electromagnetic properties for neutrinos. From a general point
of view, in 2HDMs, the symmetry breaking $SU(2)_{L}\times U(1)_{Y}\to
U(1)_{Q}$ is implemented by introducing a new scalar doublet with the same
quantum numbers of the first one. By counting the new degrees of freedom,
2HDMs contain five Higgs bosons in its spectrum \cite{HHaber}. In a
CP-conserving scenario, the Higgs sector consists of Two Higgs CP-even
scalars $\left( H^{0},h^{0}\right) ,$ one CP-odd scalar $\left( A^{0}\right) 
$ and two charged Higgs bosons $\left( H^{\pm }\right) $. A key parameter of
the model is the ratio between the vacuum expectation values, $\tan \beta ={%
v_{2}}/{v_{1}}$, where $v_{1}$ and $v_{2}$ are the VEV's of the Higgs
doublets; with values of $0\leq \beta \leq {\pi }/{2}$.

There are several ways to incorporate neutrino masses within the SM or its
extensions, to explain the observed neutrino oscillations. We shall use a
simple form which consists of adding $3$ right-handed singlets of neutrinos
fields $\left( \nu _{jR}\right) $ corresponding to each charged lepton,
enforcing the conservation of lepton number on the Lagrangian. The most
general gauge invariant Lagrangian that couples the Higgs fields to leptons
reads

\begin{equation}  \label{YukawaLagrangianI}
-\mathcal{L}_{Y}=\eta _{i,j}^{\nu,0}\overline{l}_{iL}^{0}\tilde{\Phi}
_{1}\nu_{jR}^{0}+\xi _{i,j}^{\nu,0}\overline{l}_{iL}^{0}\tilde{\Phi}%
_{2}\nu_{jR}^{0}+\eta _{i,j}^{E,0}\overline{l}_{iL}^{0}\Phi
_{1}E_{jR}^{0}+\xi _{i,j}^{E,0}\overline{l}_{iL}^{0}\Phi _{2}E_{jR}^{0}+h.c.,
\end{equation}%
where $\Phi _{1,2}\;$represents the Higgs doublets, and$\ \widetilde{\Phi }%
_{1,2}\equiv i\sigma _{2}\Phi _{1,2}$, the superscript \textquotedblleft $0$%
\textquotedblright\ indicates that the fields are not mass eigenstates yet,$%
\;\eta _{i,j}$ and $\xi _{i,j}$are non diagonal $3\times 3$ matrices with $%
\left( i,j\right);$ denoting family indices. $E_{jR}^{0}$ denotes the three
charged leptons and$\ \overline{l}_{iL}^{0};$ denotes the lepton weak
isospin left-handed doublets. In (\ref{YukawaLagrangianI}) we have
introduced non natural masses for Dirac neutrinos by considering singlets of
right handed neutrinos $\nu_{jR}$.

As was pointed before, it is customary to implement a discrete symmetry in
the 2HDM to suppress some processes such as the Flavor Changing Neutral
Currents (FCNC). In particular by demanding the $Z_{2}$ symmetry 
\begin{eqnarray}
\Phi _{1} &\rightarrow &\Phi _{1}\ \ ;\ \ \ \Phi _{2}\rightarrow -\Phi _{2};
\notag \\
E_{jR} &\rightarrow &\mp E_{jR}\ \ ;\ \ \nu_{jR}\rightarrow -\nu_{jR} ,
\label{discrete sym}
\end{eqnarray}%
FCNCs are eliminated at the tree-level. Here $\nu_{jR}$ and $E_{jR}$ denote
right-handed singlets of the down and up types of leptons respectively.

\subsection{The type I-2HDM}

By taking $E_{jR}\rightarrow -E_{jR}$, the Lagrangian (\ref%
{YukawaLagrangianI}) is reduced to the so-called type I-2HDM. In this
scenario, only $\Phi _{2}$ couples in the Yukawa sector and gives masses to
all fermions. The lepton part of the Yukawa Lagrangian in this case becomes

\begin{equation}
-\mathcal{L}_{Y}\left( \text{type I}\right) =\eta _{ij}^{E,0}\overline{l}%
_{iL}^{0}\widetilde{\Phi }_{2}\nu _{jR}^{0}+\xi _{ij}^{E,0}\overline{l}%
_{iL}^{0}\Phi _{2}E_{jR}^{0}+h.c.,
\end{equation}%
and the contribution to the lepton sector coupling with $H^{+}$ yields%
\begin{equation}
-\mathcal{L}_{Y}\left( \text{type I}\right) =\frac{g\cot \beta }{\sqrt{2}%
M_{W}}\overline{\nu }\left( U_{PMNS}M_{l}^{diag}P_{R}-M_{\nu
}^{diag}U_{PMNS}P_{L}\right) l~H^{+}+h.c.,  \label{type I expand}
\end{equation}%
where $P_{R,L}\equiv \left( 1\pm \gamma ^{5}\right) /2$. Therefore, in a
convenient chiral basis, vertices with $H^{+}$ have the following behavior 
\begin{equation}
\left( aP_{L}+bP_{R}\right) =\frac{g\cot \beta }{\sqrt{2}M_{W}}\overline{\nu 
}\left( U_{PMNS}M_{l}^{diag}P_{R}-M_{\nu }^{diag}U_{PMNS}P_{L}\right)
l~H^{+}.
\end{equation}

Working uniquely with the leptonic part, with the same couplings in the
charged sector for type I-2HDM, is possible to describe another model
without FCNCs, the so-called Flipped model. In the last scenario, leptons
and up type quarks are coupled to $\Phi_{2}$, and $\Phi_{1}$ is just coupled
to down type quarks; making type I- and Flipped models share the same
couplings between leptons and the charged-Higgs \cite{Sher}.

\subsection{The type II-2HDM}

If we use $E_{jR}\rightarrow E_{jR}\;$\ we obtain the so-called 2HDM of type
II. In this model $\Phi _{1}$ couples and gives masses to the charged lepton
sector, while $\Phi _{2}$ couples and gives masses to the neutrino sector.
Consequently, the lepton Yukawa Lagrangian becomes%
\begin{equation}
-\mathcal{L}_{Y}=\eta _{ij}^{D,0}\overline{l}_{iL}^{0}\Phi
_{1}E_{jR}^{0}+\xi _{ij}^{\nu ,0}\overline{l}_{iL}^{0}\tilde{\Phi}_{2}\nu
_{jR}^{0}+h.c.,  \label{type II}
\end{equation}%
and the term of charged current of the Lagrangian with leptons gives%
\begin{equation*}
-\mathcal{L}_{Y}\left( \text{type II}\right) =\frac{g}{\sqrt{2}M_{W}}%
\overline{\nu }\left[ \left( \cot \beta ~M_{\nu }^{diag}U_{PMNS}P_{L}+\tan
\beta ~U_{PMNS}M_{l}^{diag}P_{R}\right) \right] l~H^{+}+h.c.
\end{equation*}%
Therefore, in an appropriate chiral basis, vertices for $H^{+}$ behave as 
\begin{equation}
\left( aP_{L}+bP_{R}\right) =\frac{g}{\sqrt{2}M_{W}}\overline{\nu }\left[
\left( \cot \beta ~M_{\nu }^{diag}U_{PMNS}P_{L}+\tan \beta
~U_{PMNS}M_{l}^{diag}P_{R}\right) \right] l~H^{+}.
\end{equation}

For the type II model, an interesting aspect is that the limits of the
parameter space ($m_{H^{+}},\tan \beta $) are very similar to those obtained
by considering the minimal supersymmetric scenario. Furthermore, since we
are concerned only in lepton sectors, couplings with $H^{\pm}$ can also be
extrapolated to the lepton specific-2HDM, where $\Phi_{2}$ couples to all
quarks, while that $\Phi_{1}$ couples to right-handed leptons \cite{Bai}.

\subsection{The type III-2HDM}

When we take into account all terms in the Lagrangian (\ref%
{YukawaLagrangianI}), both doublets are coupled simultaneously to charged
leptons and neutrinos

\begin{equation}
-\mathcal{L}_{Y}=\tilde{\eta}_{i,j}^{E,0}\bar{L}_{iL}^{0}\Phi _{1}^{\prime
}E_{jR}^{0}+\tilde{\xi}_{i,j}^{E,0}\bar{L}_{iL}^{0}\Phi _{2}^{\prime
}E_{jR}^{0}+\tilde{\eta}_{i,j}^{\nu ,0}\bar{L}_{iL}^{0}\tilde{\Phi}%
_{1}^{\prime }\nu _{jR}^{0}+\tilde{\xi}_{i,j}^{\nu ,0}\bar{L}_{iL}^{0}\tilde{%
\Phi}_{2}^{\prime }\nu _{jR}^{0}+h.c.  \label{Lagrangian_type3_lep2}
\end{equation}%
Choosing a basis where Yukawa couplings $\eta _{i}$ with $\Phi _{1}$ lead to
fermion masses

\begin{eqnarray}
\eta ^{\nu } &=&\frac{\sqrt{2}}{v}M^{\nu }=\tilde{\eta}^{\nu }\cos \beta +%
\tilde{\xi}^{\nu }e^{-i\upsilon }\sin \beta, \\
\eta ^{E} &=&\frac{\sqrt{2}}{v}M^{E}=\tilde{\eta}^{E}\cos \beta +\tilde{\xi}%
^{E}e^{i\upsilon }\sin \beta.
\end{eqnarray}

Meanwhile, Yukawa couplings $\xi_{i}$ with $\Phi_{e}$ lead to FCNC couplings 
\begin{eqnarray}
\xi ^{\nu } &=&-\tilde{\eta}^{\nu }\sin \beta +\tilde{\xi}^{\nu
}e^{-i\upsilon }\cos \beta, \\
\xi ^{E} &=&-\tilde{\eta}^{E}\sin \beta +\tilde{\xi}^{E}e^{i\upsilon }\cos
\beta .
\end{eqnarray}
By a biunitary transformation involving $V_{L}^{\nu },V_{R}^{\nu },V_{L}^{E}$
and $V_{R}^{E}$ matrices, Yukawa couplings can be expressed in the basis of
lepton masses, where $\bar{\eta}^{\nu }$ and $\bar{\eta}^{E}$ mass matrices
will be diagonal and real

\begin{eqnarray}
\bar{\eta}^{\nu } &=&\frac{\sqrt{2}}{v}M^{\nu }=V_{L}^{\nu }\eta ^{\nu
}V_{R}^{\nu \dagger },\text{ \ }M_{ij}^{\nu }=\delta _{ij}m_{i}^{\nu }; \\
\bar{\eta}^{E} &=&\frac{\sqrt{2}}{v}M^{E}=V_{L}^{E}\eta ^{E}V_{R}^{E\dagger
},\text{ \ }M_{ij}^{D}=\delta _{ij}m_{i}^{E},
\end{eqnarray}

and the remaining couplings are associated to FCNC matrices

\begin{eqnarray}
\bar{\xi}^{\nu } &=&V_{L}^{\nu }\xi ^{\nu }V_{R}^{\nu \dagger } \\
\bar{\xi}^{E} &=&V_{L}^{D}\xi ^{E}V_{R}^{E\dagger }.
\end{eqnarray}

To write Lagrangian in Eq. (\ref{Lagrangian_type3_lep2}) in terms of mass
eigenstates, we should make a unitary transformation over singlets

\begin{eqnarray}
E_{L,R} &=&V_{L,R}^{E}E_{L,R}^{0}, \\
\nu _{L,R} &=&V_{L,R}^{\nu }\nu _{L,R}^{0}.
\end{eqnarray}

Performing a biunitary transformation, FCNC Lagrangian can be written by

\begin{equation}
\mathcal{L}_{FCNC}=\bar{\xi}_{i,j}^{\nu }\bar{L}_{iL}\tilde{H}_{2}\nu _{jR}+%
\bar{\xi}_{i,j}^{E}\bar{L}_{iL}H_{2}E_{jR}+h.c.,
\end{equation}%
where $V_{R}^{\nu ,E}$ is completely unknown and FCNC matrices $\xi ^{E,\nu
} $ are arbitrary. Under specific conditions, we should make a
parametrization with a hierarchical structure in an analog way for fermion
masses. It is achieved by imposing a convenient texture for nondiagonal
matrices \cite{Sher}, i.e. proposing that FCNC couplings should be of the
order of the geometric mean of the masses

\begin{equation*}
\xi _{ij}=\lambda _{ij}\frac{\sqrt{2m_{i}m_{j}}}{v},
\end{equation*}%
where $\lambda _{ij}$ are of $\mathcal{O}(1)$. This condition is the
so-called Sher-Cheng anzats. In the lepton sector, the matrix texture taken
would have the following structure \cite{Ahuja, Diaz-cruz,Hernandez-S}.

\begin{eqnarray}
\xi ^{E} &=&\left( 
\begin{array}{ccc}
0 & \times & 0 \\ 
\times & 0 & \times \\ 
0 & \times & \times%
\end{array}%
\right) ^{l}=\frac{\sqrt{2}\lambda _{l}}{v}\left( 
\begin{array}{ccc}
0 & \sqrt{m_{e}m_{\mu }} & 0 \\ 
\sqrt{m_{e}m_{\mu }} & 0 & \sqrt{m_{\tau }m_{\mu }} \\ 
0 & \sqrt{m_{\tau }m_{\mu }} & m_{\tau }%
\end{array}%
\right); \\
\xi ^{\nu } &=&\left( 
\begin{array}{ccc}
0 & \times & 0 \\ 
\times & 0 & \times \\ 
0 & \times & \times%
\end{array}%
\right) ^{\nu }=\frac{\sqrt{2}\lambda _{\nu }}{v}\left( 
\begin{array}{ccc}
0 & \sqrt{m_{\nu _{e}}m_{\nu _{\mu }}} & 0 \\ 
\sqrt{m_{\nu _{e}}m_{\nu _{\mu }}} & 0 & \sqrt{m_{\nu _{\tau }}m_{\nu _{\mu
}}} \\ 
0 & \sqrt{m_{\nu _{\tau }}m_{\nu _{\mu }}} & m_{\nu _{\tau }}%
\end{array}%
\right) ,  \label{textura_2hdm}
\end{eqnarray}

where we have made the extrapolation from charged lepton part to neutrino
sector, by assuming the same mass hierarchy as in the charged lepton sector.
Furthermore, textures have the assumption that $\lambda^{\prime }s$ acting
on the mass matrix are the same for all FCNC couplings.

In the Higgs basis, interactions among charged leptons $l$ and neutrinos $%
\nu _{l}$ arise from the following Lagrangian%
\begin{equation}
-\mathcal{L}_{Y}\left( \text{type III}\right) =\bar{\nu}_{l}\left(
U_{PMNS}\xi ^{E}P_{R}-\xi ^{\nu }U_{PMNS}P_{L}\right) lH^{+}+h.c.
\end{equation}

Under any assumption in the couplings in the type III-2HDM, the radical
difference with other 2HDMs presenting FCNC suppression is that the magnetic
dipole moment arises as a consequence of flavor neutral currents, but not
directly from mass terms in the Lagrangian.

\section{Neutrino masses in neutrino specific-2HDM\label{nspecific}}

We have considered in this paper an extension of the Standard Model
consisting of adding one extra Higgs doublet with the same quantum numbers
as those for SM-doublet along with three right-handed neutrinos, which are
singlets under the Standard Model $SU(2)\times $ $U(1)$ gauge group. The
VEV's for both doublets belong in the same scale of energy. Neutrinos couple
to doublets at the same energy scale that charged leptons yielding a
hierarchical problem for fermions masses. Thus, in those scenarios and to
take into account Dirac masses and explain MDMs, right-handed neutrinos are
implemented in an unnatural way.

We shall work on the so-called Neutrino specific-2HDM, where the tiny
neutrino masses could arise using a small VEV, which implies fewer
assumptions than fitting small masses or Yukawa couplings directly. In this
framework, we intend to calculate electromagnetic form factors for
neutrinos. For this model, it is customary to implement a $U\left( 1\right) $
global symmetry defined by 
\begin{equation}
\Phi _{1}\rightarrow e^{i\varphi }\Phi _{1}\text{ and }\Phi _{2}\rightarrow
-\Phi _{2}.
\end{equation}

The most general Lagrangian in neutrino specific model reads%
\begin{equation}
\mathcal{L}=\left( \mathcal{D}_{\mu }\Phi _{1}\right) ^{\dagger }\left( 
\mathcal{D}^{\mu }\Phi _{1}\right) +\left( \mathcal{D}_{\mu }\Phi
_{2}\right) ^{\dagger }\left( \mathcal{D}^{\mu }\Phi _{2}\right) -V_{H}+%
\mathcal{L}_{Y}+\mathcal{L}_{\nu },  \label{neutrinospecific}
\end{equation}

where $V_{H},$ $\mathcal{L}_{Y}$ and $\mathcal{L}_{\nu }$ denote the Higgs
potential, Yukawa Lagrangian for quarks and charged leptons and Yukawa
Lagrangian for the neutrino sector respectively. One interesting and
straightforward model arising from this Lagrangian structure is the neutrino
specific 2HDM, wherein Dirac neutrino masses are generated with the same SSB
to the remaining fermions. Naturalness and smallness are circumvented by the
fact that a doublet acquires one VEV in the same scale as the neutrino
masses \cite{Gabriel,logan1}. To see this formally, we introduce a Higgs
potential taking into account a softly broken $U\left( 1\right) $ symmetry
using a $\bar{m}_{12}^{2}$ \ term, which has small radiative corrections as
well as soft contributions to RGE's \cite{logan2}

\begin{align}
V_{H}& =\bar{m}_{11}^{2}\Phi _{1}^{\dagger }\Phi _{1}+\bar{m}_{22}\Phi
_{2}^{\dagger }\Phi _{2}-\left( \bar{m}_{12}^{2}\Phi _{1}^{\dagger }\Phi
_{2}+h.c.\right)  \notag \\
& +\frac{1}{2}\lambda _{1}\left( \Phi _{1}^{\dagger }\Phi _{1}\right) ^{2}+%
\frac{1}{2}\lambda _{2}\left( \Phi _{2}^{\dagger }\Phi _{2}\right)
^{2}+\lambda _{3}\left( \Phi _{1}^{\dagger }\Phi _{1}\right) \left( \Phi
_{2}^{\dagger }\Phi _{2}\right) +\lambda _{4}\left( \Phi _{1}^{\dagger }\Phi
_{2}\right) \left( \Phi _{2}^{\dagger }\Phi _{1}\right) .
\label{HiggsPotential1}
\end{align}

We can settle $\bar{m}_{12}^{2}$ as real and positive without loss of
generality, by re-phasing $\Phi _{2}$ and translating the excess phase into
the Yukawa couplings, which is already, in general, a complex matrix taking
into account possible CP violation phases. Note that even after the global $%
U(1)$ is softly broken, lepton number survives as an accidental symmetry of
the model. Neutrinoless double beta decay and Majorana neutrinos are thus
absent.

On the other hand, it is worthwhile to discuss naturalness dependency in the
scalar spectrum and radiative corrections for propagators. Likewise SM, the
mass-squared parameters $\bar{m}_{11}^{2}$ and $\bar{m}_{22}^{2}$ in the
Higgs potential (\ref{HiggsPotential1}) suffer from large radiative
corrections with quadratic $\Lambda ^{2}$ sensitivity to the high-scale
cutoff of the theory, in the same way as the SM Higgs mass-squared parameter
works. Hence, hierarchy problem is still present in this model. Nonetheless,
it is possible to avoid it, by introducing this scalar sector in a model
with strong dynamics (e.g. Composite Higgs Model) or SUSY in TeV scale. Due
to smallness size demanded for $\bar{m}_{12}^{2}$\footnote{%
In this $U\left( 1\right) $ softly broken model $\bar{m}_{12}^{2}$ has the
following RGE%
\begin{equation*}
16\pi ^{2}\frac{d}{d\log \mu ^{2}}\bar{m}_{12}^{2}=\left( 2\lambda
_{3}+\lambda _{4}\right) \bar{m}_{12}^{2}.
\end{equation*}%
Hence its size is technically natural since radiative corrections to $%
m_{12}^{2}$ are proportional to $m_{12}^{2}$ itself and are only
logarithmically sensitive to the cutoff.}, this problem can be circumvented
even a TeV scale when mass scalar eigenstates $H^{0}$ and $A^{0}$ take
appropriate values in the mass range of $400-600$ GeV \cite{Barbieri}.

Mass eigenstates in this model can be achieved by the doublets
parameterization 
\begin{equation}
\Phi _{i}=\frac{1}{\sqrt{2}}\left( 
\begin{array}{c}
\sqrt{2}\phi _{i}^{+} \\ 
\phi _{i}^{0}+v_{i}+i\xi _{i}%
\end{array}%
\right).
\end{equation}

Relations among quartic couplings and Higgs masses are \footnote{%
By perturbativity grounds, we expect that small differences between $%
m_{A^{0}}^{2}$ and $m_{H^{0}}^{2}$ must be at most of the same order of $%
v_{2}^{2}.$}

\begin{eqnarray}
m_{h^{0}}^{2} &=&\lambda _{1}v^{2},  \notag \\
m_{H^{\pm }}^{2} &=&m_{22}^{2}+\frac{1}{2}\lambda _{3}v^{2}, \\
m_{A,H}^{2} &=&m_{H^{\pm }}^{2}+\frac{1}{2}\lambda _{4}v^{2}.  \notag
\end{eqnarray}

This model behaves as a pseudo-inert 2HDM, where $v_{2}<<v_{1}$. The Yukawa
Lagrangian in (\ref{neutrinospecific}) with massive neutrinos (considering
right-handed singlets of neutrinos) inspired on 2HDM reads

\begin{equation}
-\mathcal{L}_{Y}=\xi _{ij}^{\nu }\bar{L}_{Li}\tilde{\Phi}_{2}\nu _{Rj}+\eta
_{ij}^{E}\bar{L}_{Li}\Phi _{1}E_{Rj}+\eta _{ij}^{D}\bar{Q}_{Li}\Phi
_{1}D_{Rj}+\eta _{ij}^{U}\bar{Q}_{Li}\tilde{\Phi}_{1}U_{Rj}+h.c.,
\label{NELagrangian}
\end{equation}

where $\tilde{\Phi}_{i}=i\sigma _{2}\Phi _{i},$ is the conjugate of the
Higgs doublet. Fermion doublets are defined by $Q_{L}\equiv \left(
u_{L},d_{L}\right) ^{T}$ and $L_{L}\equiv \left( \nu _{L},e_{L}\right) ^{T}.$
Here $E_{R}\left( D_{R}\right) $ is referred to the three down type weak
isospin lepton(quark) singlets and $\nu _{R}\left( U_{R}\right) $ is
referred to the three up type weak isospin neutrino (quark) singlets. The
first part in (\ref{NELagrangian}) comes from right-handed neutrinos, while
the second part is SM like. This can be considered as a Dirac addition for
SM, where the small mass for neutrinos come from a small VEV in the second
doublet. These right handed neutrinos $\nu _{R}$ will pair up with the three
left-handed neutrinos of the SM to form Dirac particles. Compatibility with
a Higgs potential lead to define a $U\left( 1\right) $ charge acting only
over the second doublet. Hence all SM fields hold unchanged. The Yukawa
Lagrangian might also be generated from a $U(1)$-global symmetry from the
following set of transformations 
\begin{eqnarray}
\Phi _{1} &\rightarrow &e^{i\varphi }\Phi _{1}\text{ and }\Phi
_{2}\rightarrow -\Phi _{2},  \notag \\
E_{jR} &\rightarrow &e^{-i\omega }E_{jR}\text{ and }\nu _{jR}\rightarrow
-\nu _{jR}. \\
D_{jR} &\rightarrow &e^{-i\omega }D_{jR}\text{ and }U_{jR}\rightarrow
e^{-i\varphi }U_{jR}.  \notag
\end{eqnarray}

with $\omega =\varphi \ $(which can be taken equal to zero as a particular
case) generates neutrino-specific 2HDM. Moreover, it is possible to see that
these transformations do not allow Majorana terms because of the presence of
only right-handed singlets, i.e., $\mathcal{L}_{Maj}=\eta ^{M}\overline{%
\left( \nu _{R}^{\ast }\right) }\tilde{\Phi}_{2}\nu _{R}.$ The lepton sector
of the Yukawa Lagrangian in its charged part reads%
\begin{equation}
-\mathcal{L}_{Y}\left( \text{neutrino specific}\right) =\frac{\sqrt{2}m_{\nu
_{i}}}{v_{2}}\bar{\nu}_{l}\left( U_{PMNS}P_{L}\right) lH^{+}+h.c.
\end{equation}

\section{Radiative corrections in 2HDM\label{loops}}

\begin{figure}[tph]
\centering
\includegraphics[scale=1]{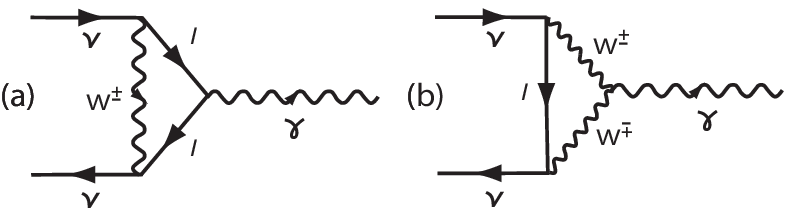} \includegraphics[scale=1]{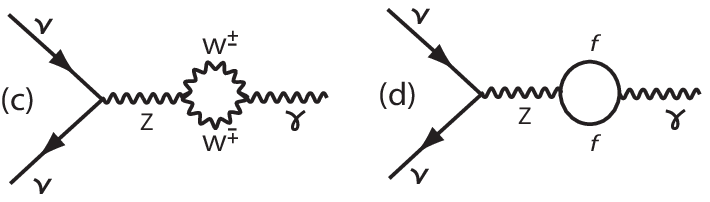} \vspace{%
-0.3cm}
\caption{\textit{Loop contributions in SM to neutrino electromagnetic form
factors.}}
\label{fig3}
\end{figure}

With all fundamentals for lepton charged Higgs sectors in different 2HDMs in
mind, we study new physics contributions to neutrino-MDM. We begin our
discussion with the background contribution in the SM with right-handed
neutrinos. Figure \ref{fig3} shows diagrams contributing to the neutrino
electromagnetic vertex in SM+RH neutrinos.

\begin{figure}[tph]
\centering
\includegraphics[scale=1]{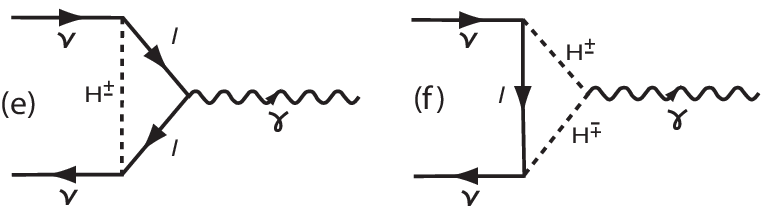} \includegraphics[scale=1]{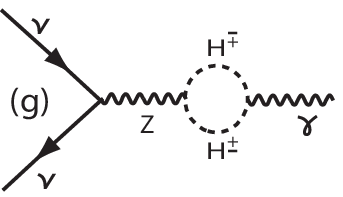} 
\vspace{-0.3cm}
\caption{\textit{Loop corrections coming from the 2HDM new physics, in which
charged Higgs bosons run into the triangle contribution.}}
\label{fig3.1}
\end{figure}

Within the framework of a 2HDM with massive neutrinos, we should add three
new types of contributions: two vertex corrections and one correction to the
vacuum polarization displayed in Fig. \ref{fig3.1}. They arise by replacing $%
W^{\pm }$ by $H^{\pm }$ in the SM diagrams

We can parametrize new physics effects by separating the contributions of SM
and 2HDM

\begin{equation}
\Lambda _{2HDM}=\Lambda _{SM}+\Delta \Lambda _{2HDM},
\end{equation}%
where $\Lambda _{SM}$ provides the contribution for MDM coming from the SM,
given by Eq. (\ref{MDMSM}). The new physics contribution $\Delta \Lambda
_{2HDM}$ splits into two diagrams: i) The first one associated with a vertex
correction with two charged Higgs bosons and one charged lepton ($\Delta
\Lambda _{2H\pm 1L}$) into the loop. ii) The second one is the vertex
correction with two charged leptons and one charged Higgs boson ($\Delta
\Lambda _{2L1H\pm }$) into the loop. Finally, the diagram in Fig. \ref%
{fig3.1}(g), does not contribute to MDM.

Figure \ref{fig3.1}(f) shows the vertex correction involving two charged
Higgs bosons and one charged lepton into the loop $\left( 2H^{\pm }1L\right) 
$. For this diagram, the general form of the contribution can be written as 
\begin{equation}
\Delta \Lambda _{2H^{\pm }1L}^{\alpha }\left( q,l\right) =-e\int \frac{d^{4}k%
}{\left( 2\pi \right) ^{4}}\left\{ \frac{\left( aP_{L}+bP_{R}\right) \left(
2k^{\alpha }+p_{2}^{\alpha }+p_{1}^{\alpha }\right) \left(
cP_{L}+dP_{R}\right) \left( \NEG{k}+m_{l}\right) }{\left[ \left(
k+p_{1}\right) ^{2}-m_{H^{\pm }}^{2}\right] \left[ \left( k+p_{2}\right)
^{2}-m_{H^{\pm }}^{2}\right] \left( k^{2}-m_{l}^{2}\right) }\right\} ,
\end{equation}

where $a,b,c$ and $d$ are constants associated with the Feynman rules of the
particular 2HDM. Finally, $\alpha $ is a Lorentz index.

On the other hand, the diagram in Fig. \ref{fig3.1}(e) with two leptons and
one charged Higgs into the loop $\left( 2L1H^{\pm }\right) $, gives a
contribution of the form

\begin{equation}
\Delta \Lambda _{2L1H^{\pm }}^{\alpha }\left( q,l\right) =-e\int \frac{d^{4}k%
}{\left( 2\pi \right) ^{4}}\left\{ \frac{\left( aP_{L}+bP_{R}\right) \left( 
\NEG{k}+\NEG{p}_{1}+m_{l}\right) \gamma ^{\alpha }\left( \NEG{k}+\NEG%
{p}_{2}+m_{l}\right) \left( cP_{L}+dP_{R}\right) }{\left[ \left(
k+p_{1}\right) ^{2}-m_{l}^{2}\right] \left[ \left( k+p_{2}\right)
^{2}-m_{l}^{2}\right] \left( k^{2}-m_{H^{\pm }}^{2}\right) }\right\} .
\end{equation}

Then by factorizing out these integrals in the form of Eq. (\ref{Fullvertex}%
), contribution from new physics to magnetic dipole moment is finally

\begin{eqnarray}
\Delta \Lambda _{2HDM}\left( q,l\right) _{F_{M}} &=&2\Delta \Lambda
_{2H1L}\left( q,l\right) _{F_{M}}+2\Delta \Lambda _{2L1H}\left( q,l\right)
_{F_{M}}  \notag \\
&=&\frac{ei}{8\pi ^{2}}\int\limits_{0}^{1}dx\int\limits_{0}^{x}dy\frac{1}{%
D_{1}}\left[ m_{\nu }\left( -1+3x-x^{2}\right) +m_{l}\left( \frac{1}{2}%
-x\right) \right] \left( ac+bd\right)   \notag \\
&&+\frac{ei}{8\pi ^{2}}\int\limits_{0}^{1}dx\int\limits_{0}^{x}dy\frac{1}{%
D_{2}}\left[ m_{\nu }\left( 4x^{2}-5x+2\right) \left( ad+bc\right)
+m_{l}x\left( ac+bd\right) \right] ,  \label{2HDM_V_FM}
\end{eqnarray}

where the number 2 comes from the inclusion of both $H^{+}$ and $H^{-}$ into
the contributions. $D_{1}$ and $D_{2}$ are defined as 
\begin{eqnarray*}
D_{1} &=&y^{2}m_{\nu }^{2}-2xym_{\nu }^{2}+\left( m_{\nu
}^{2}-m_{l}^{2}+m_{H^{\pm }}^{2}\right) y+x^{2}m_{\nu }^{2}+\left(
m_{l}^{2}-m_{\nu }^{2}-m_{H^{\pm }}^{2}\right) x+m_{H^{\pm }}^{2}, \\
D_{2} &=&y^{2}m_{\nu }^{2}-\left( m_{H^{\pm }}^{2}-m_{\nu
}^{2}-m_{l}^{2}\right) y-2xym_{\nu }^{2}+x^{2}m_{\nu }^{2}-\left(
m_{l}^{2}+m_{\nu }^{2}-m_{H^{\pm }}^{2}\right) x+m_{l}^{2}.
\end{eqnarray*}

Using Yukawa Lagrangians for charged scalar sector described in sections \ref%
{2HDM} and \ref{nspecific}, it is possible to extract the contributions for
different models to charged Higgs-charged lepton and neutrino vertices,
which are summarized in Tab. \ref{tab:couplingsPLPR}. 
\begin{table}[tph]
\centering%
\begin{tabular}{|l|c|c|c|c|}
\hline
Vertex couplings & Type I and Flipped & Type II and Lepton specific & Type
III & Neutrino specific \\ \hline\hline
$a=c$ & $-2^{\frac{3}{4}}\sqrt{G_{F}}m_{v_{l}}\cot \beta U_{k,i}$ & $2^{%
\frac{3}{4}}\sqrt{G_{F}}m_{v_{l}}\cot \beta U_{k,i}$ & $-\xi _{i,k}^{\nu
}U_{k,i}$ & $-\frac{\sqrt{2}}{v_{2}}m_{v_{i}}U_{k,i}$ \\ \hline
$b=d$ & $2^{\frac{3}{4}}\sqrt{G_{F}}m_{l}\cot \beta U_{i,k}$ & $2^{\frac{3}{4%
}}\sqrt{G_{F}}m_{l}\tan \beta U_{i,k}$ & $U_{i,k}\xi _{k,i}^{E}$ & $\frac{%
\sqrt{2}}{v_{2}}m_{v_{i}}U_{i,k}$ \\ \hline
\end{tabular}%
\caption{\textit{Coefficients for $P_{L}$ and $P_{R}$ couplings present in
the Magnetic Dipole Moment for type I,II,III and neutrino specific-2HDMs. In
our numerical analyses, we shall use the accurate value for Fermi's constant 
$G_{F}=\protect\sqrt{2}g^{2}/8 M_{W}^{2}=1.1663787(6)\times 10^{-5}\hspace{%
0.1cm}\text{GeV}^{-2}$. }}
\label{tab:couplingsPLPR}
\end{table}

\section{Results and analysis for MDM in 2HDMs\label{results}}

Our analysis are based on constraints on charged Higgs masses. For either
type I or II models, the experimental constraints on the possible values in
the $\left( m_{H^{\pm }},\tan \beta \right) $ parameter space come from
processes such as $B_{u}\rightarrow \tau \nu _{\tau },~D_{s}\rightarrow \tau
\nu _{\tau },~B\rightarrow D\tau \nu _{\tau },~K\rightarrow \mu \nu _{\mu }$
and $\text{BR}\left( B\rightarrow X_{s}\gamma \right) $ \cite{Akeroyda}.

Using the phenomenological constraints on the type I-2HDM , we take values
of $\tan \beta $ between $\left( 2-90\right) $ and values of the charged
Higgs mass of $m_{H^{\pm }}=\left(100-900\right) $ GeV \cite{Mahmoudi}. On
the other hand, for the type II-2HDM, we have different allowed intervals of 
$\tan \beta $ for different values of the charged Higgs mass: for $m_{H^{\pm
}}=300$ GeV the values of $\tan \beta $ lie within the interval $\left(
4-40\right) $, for $m_{H^{\pm}}=400$ GeV, $\tan\beta$ has the allowed
interval of $\left(2-55\right)$. For $m_{H^{\pm }}=500$ GeV the value of $%
\tan \beta $ is between $\left( 2-69\right) $ and for $m_{H^{\pm }}=\left(
700-900\right) $ GeV, $\tan \beta $ is between $\left( 1-70\right) \ $ \cite%
{Sher, Mahmoudi}.

\begin{figure}[tph]
\centering
\textbf{{\small Type I (Flipped) and II (Lepton-specific) \hspace{2.0cm}
Type III (Sher and Cheng Anzats)}}\newline
\includegraphics[width=0.48\columnwidth]{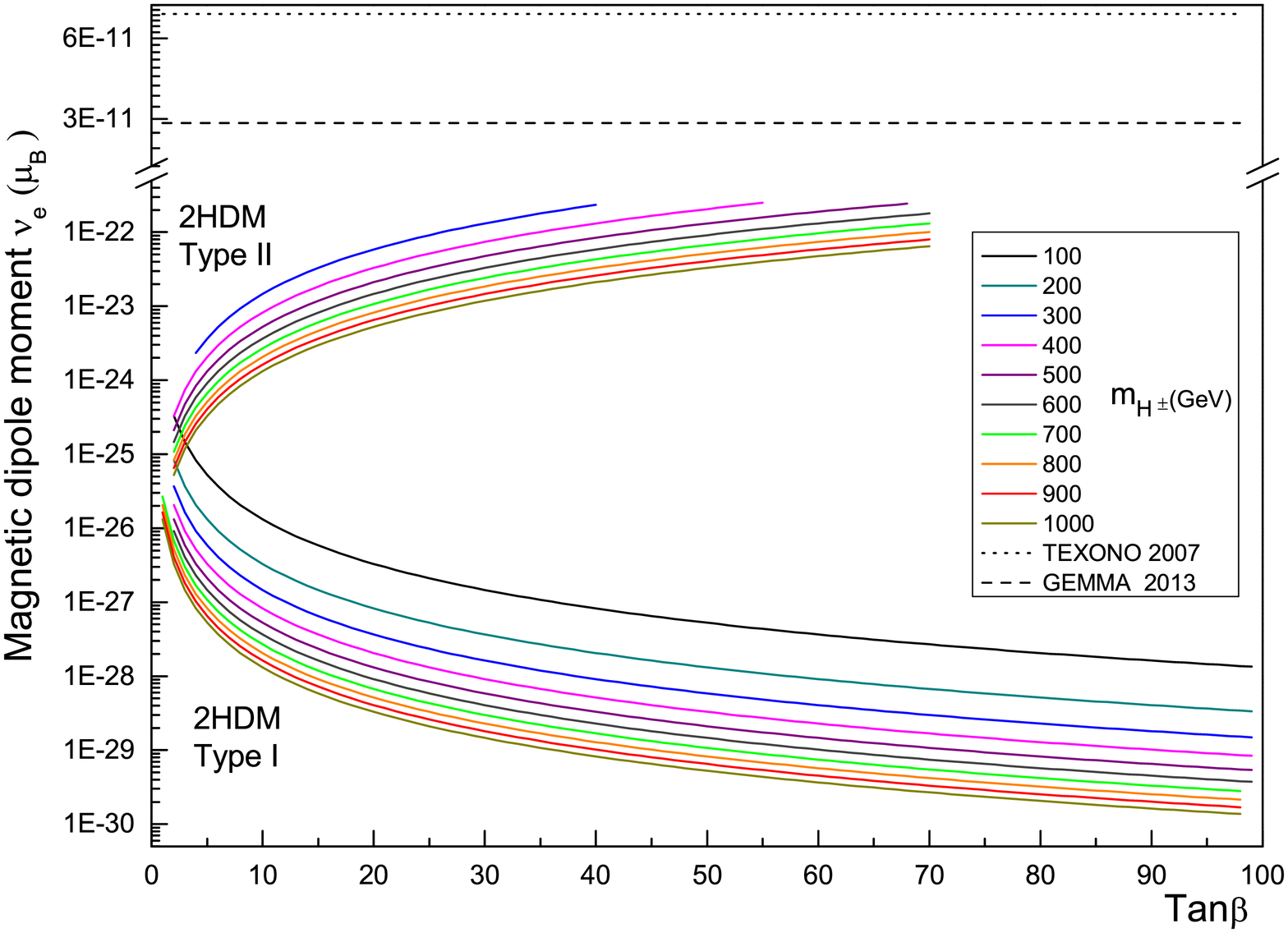} %
\includegraphics[width=0.48\columnwidth]{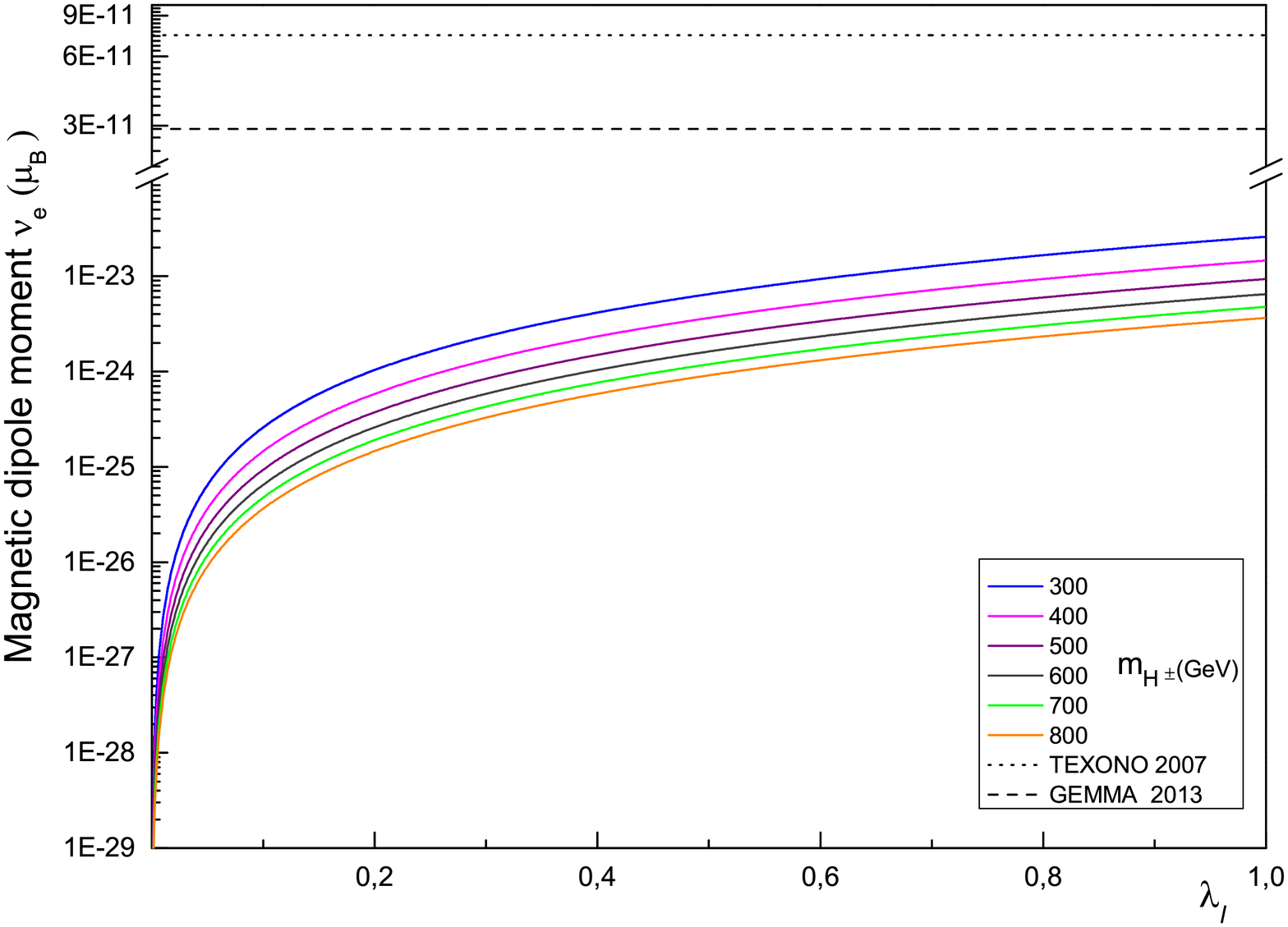} \vspace{0.3cm}
\includegraphics[width=0.48\columnwidth]{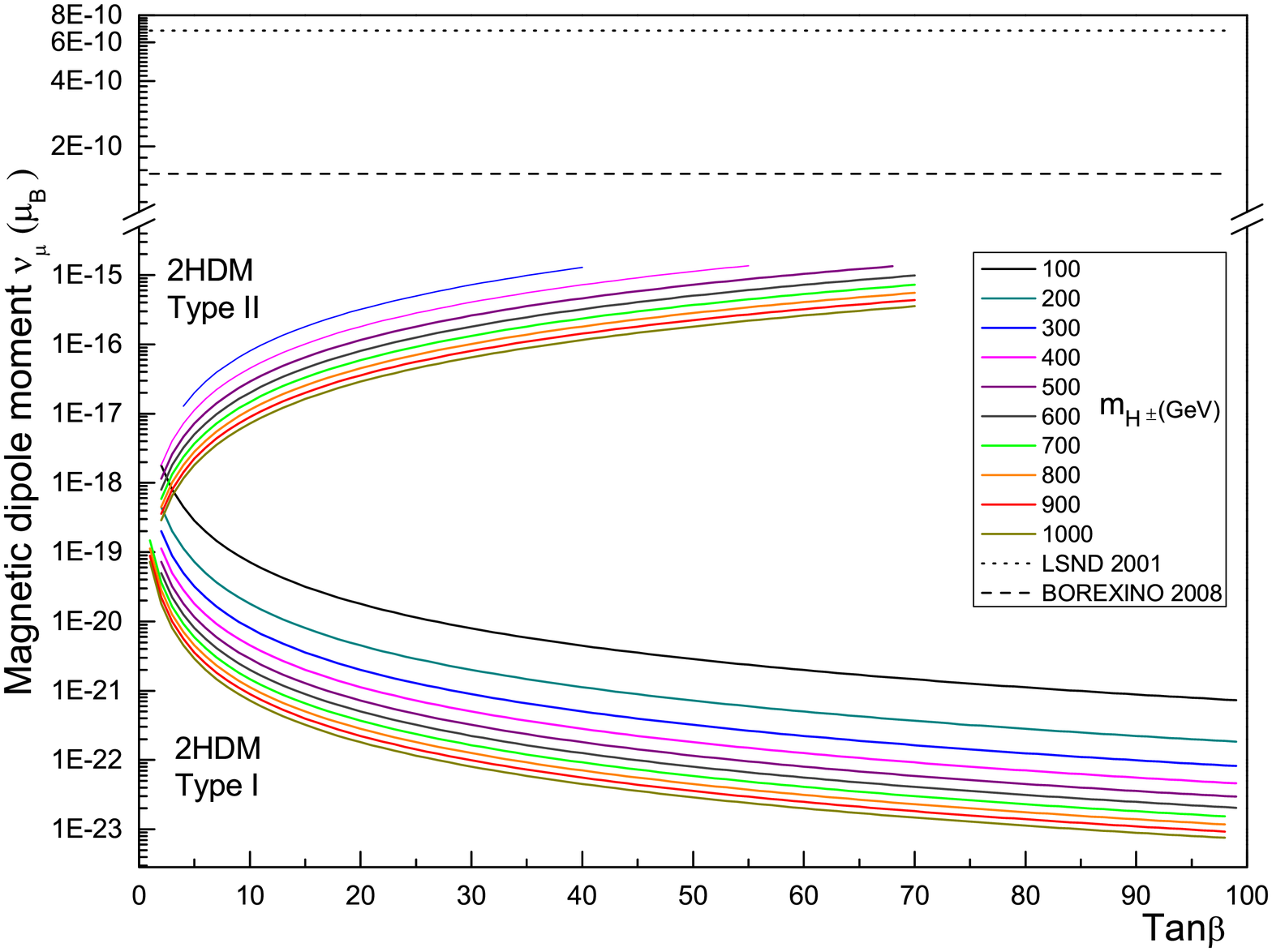} %
\includegraphics[width=0.48\columnwidth]{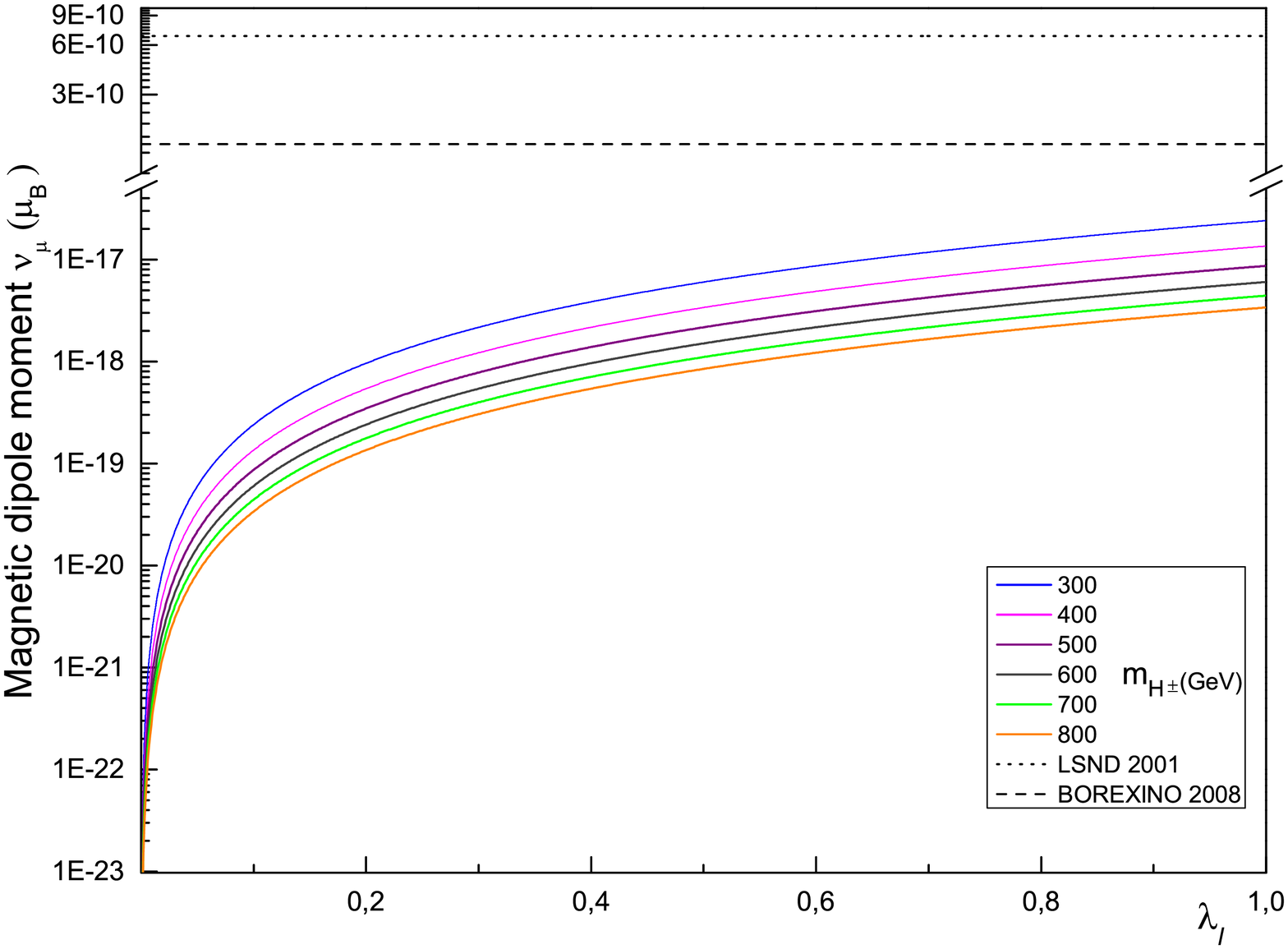} \vspace{0.3cm} %
\includegraphics[width=0.48\columnwidth]{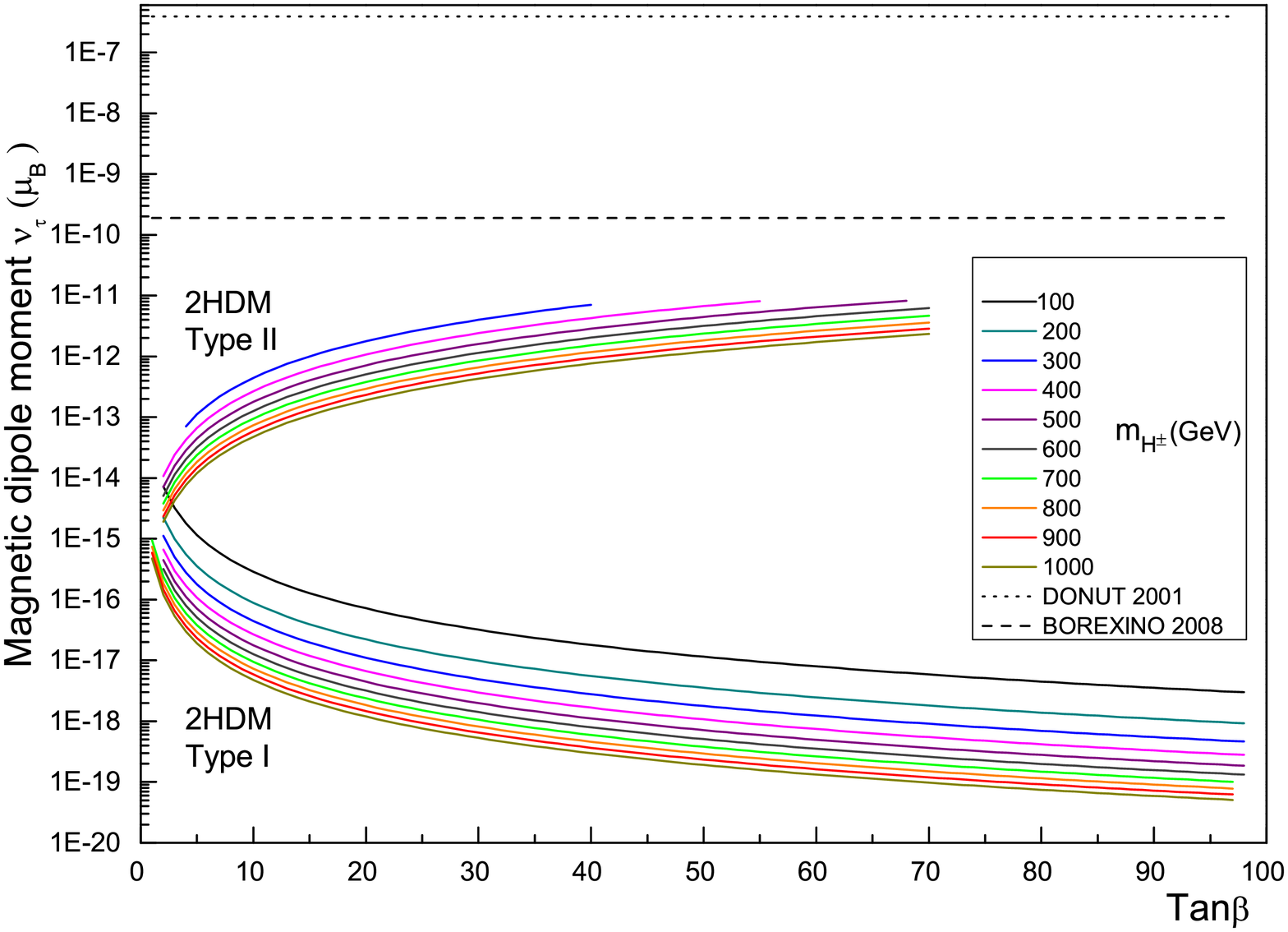} %
\includegraphics[width=0.48\columnwidth]{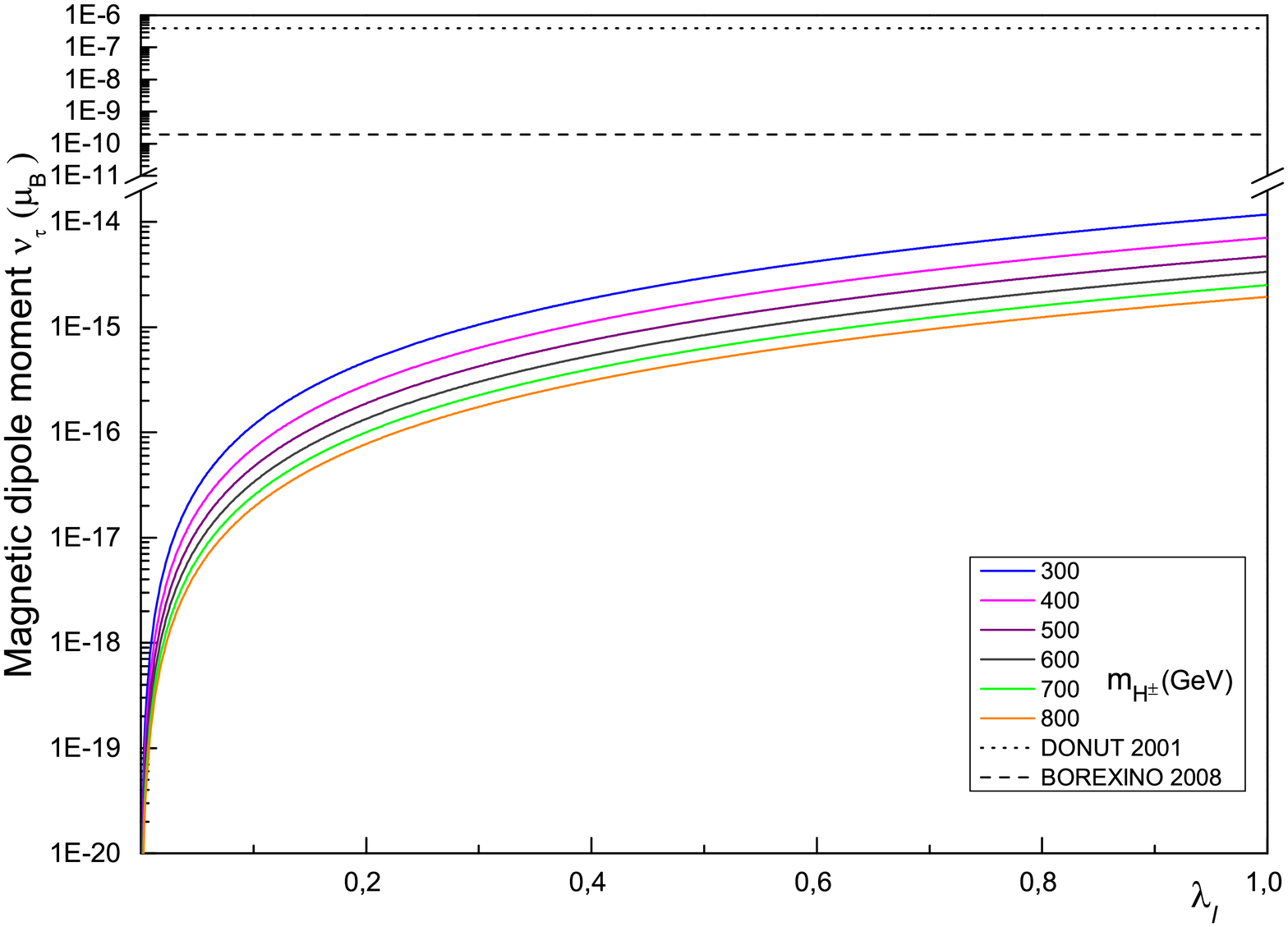} \vspace{-0.4cm}
\caption{\textit{(\textbf{Left}) Contribution to the magnetic dipolar moment
for $\protect\nu_{e},\protect\nu_{\protect\mu},\protect\nu_{\protect\tau}$%
-neutrinos coming from type I (and Flipped), II ( and Lepton-specific) 2HDMs
with masses of charged Higgs sweeping between }$(100-900)$\textit{\ GeV to
type I, }$(300-900)$\textit{\ GeV to type II to different values of }$\tan 
\protect\beta $\textit{\ to each mass of charged Higgs. (\textbf{Right})
Contribution to the magnetic dipolar moment for neutrinos coming from type
III-2HDM. Here we have taken the masses of charged Higgs sweeping between }$%
(300-800)$ and $\protect\lambda _{\protect\nu },\protect\lambda _{l}\in %
\left[ 10^{-6},1\right] $. \textit{The horizontal dotted line makes
reference to the experimental thresholds for each neutrino flavor at $90\%$
C.L.. }}
\label{type_I_II}
\end{figure}

In the first three graphics of Fig. \ref{type_I_II}, we plot the electron,
muon and tau neutrino MDM versus $\tan \beta $ for the same charged Higgs
masses as before for the type I and II 2HDMs. The horizontal lines in the
electron neutrino case corresponds to the experimental upper limits for MDM
coming from TEXONO 2007 (Taiwan EXperiment On NeutriNO) \cite{Wong} which is 
$\mu _{\overline{\nu }_{e}}<7.4\times 10^{-11}\mu _{B}$ at $90\%~$ C.L., and
GEMMA 2013. (Germanium Experiment for measurement of Magnetic Moment of
Antineutrino) \cite{Gemma} which is $\mu _{\bar{\nu}_{e}}<2.9\times
10^{-11}\mu _{B}$ at $90\%~$ C.L.. In the case of muon neutrino the
horizontal lines correspond to the experimental limits for MDM coming from
LSND 2001(Liquid Scintillating Neutrino Detector) \cite{Auerbach} that
yields $\mu _{\nu _{\mu }}<6.8\times 10^{-10}\mu _{B}$ at $90\%~$ C.L., and
BOREXino 2008 BOREXino is the Italian diminutive of BOREX (Boron solar
neutrino experiment) \cite{Montanino} that gives $\mu _{\nu _{\mu
}}<1.9\times 10^{-10}\mu _{B}$ at $90\%~$ C.L. Finally the horizontal lines
to tau neutrino correspond to the experimental limits for MDM coming from
DONUT 2001(Direct Observation of the NU Tau) \cite{Schwienhorst} that gives $%
\mu _{\nu _{\tau }}<3.9\times 10^{-7}\mu _{B}$ at $90\%~$ C.L., and BOREXino
2008 \cite{Montanino} whose upper limit is $\mu _{\nu _{\tau }}<1.5\times
10^{-10}\mu _{B}$ at $90\%~$ C.L.\footnote{%
In a current study \cite{ValleNew}, an updated analysis of the neutrino
magnetic moments was provided. Particularly, based on the most recent data
from Borexino, a new limit on the effective neutrino magnetic moment has
been obtained: $3.1\times 10^{-11}$ $\mu _{B}$ at $90\%$ C.L.}.

We analyze magnetic dipolar moments for 2HDMs, taking the values for
effective masses of each flavor neutrino considered in Tab \ref{tab:EMNC}
(which are respecting the cosmological bound and mass differences for $%
\nu_{1},\nu_{2}$ and $\nu_{3}$ states in a normal ordering):

\begin{itemize}
\item Magnetic dipole moment of $\nu _{e}$: For the type I and Flipped
cases, the largest contributions come from lower values of $\tan \beta <10$,
that provides a scale for the MDM of at least four magnitude orders between $%
10^{-28}-10^{-25}$ $\mu _{B}$; being the most relevant for low values of
charged Higgs masses. Hence, new physics from the type I-2HDM does not give
a significant correction on SM effective operators ($\Lambda _{SM}\sim
2\times 10^{-20}\mu _{B}$). Therefore, there are no regions of exclusions
for the parameter space of type I from MDM's analyses. As for the type II
case, the most significant contributions are established by lower values of
charged Higgs mass and higher values of $\tan \beta $; our better case $\tan\beta =55$ and
a mass of charged Higgs of 400 GeV with a contribution close to $2\times10^{-22}$ $\mu _{B}$. 
Even though this contribution is
higher than the one of the type I-2HDM, they are still around of two
orders of magnitude below of SM contribution.

Therefore, differences between both models start in one order of magnitude
(for$\ 1<\tan \beta <10$) and go up to at least seven orders of magnitude
(for$\ \tan \beta >>1$). The difference in behavior between both models with
respect to$\ \tan \beta $ is a consequence of couplings structure in the
Yukawa sectors.

In the type III-2HDM, relevant contributions are located for $%
\lambda^{\prime }s\to 1$ and lower values of charged Higgs mass. For these
values, the maximum achieved for $\nu_{e}$-MDM is three orders of magnitude
below of SM contribution.

\item Magnetic dipole moment of $\nu _{\mu }$: Concerning MDM for electron
neutrino, the value for a similar parameter space of new physics is at a
higher scale, which lies between five to six orders of magnitude above.
Fundamentally, these effects are slightly due to our assumption of a normal
hierarchy for neutrinos. Despite contributions of the type II-2HDM overcome
SM values, our better case $\tan\beta =55$ and
a mass of charged Higgs of 400 GeV with a contribution close to $1\times10^{-15}$ $\mu_{B}$, 
avoiding possible exclusion regions in
the parameter space of these theories. A similar case is presented in the
type III-scenario (using Cheng-Sher anzats), where the best contribution is
close to $10^{-17}$ $\mu_{B}$ ($m_{H^{\pm}}=300$ Gev and $\lambda_{l}\to 1$).

\item Magnetic dipole moment of $\nu _{\tau }$: We can see that the scale
for MDM increases by almost four orders of magnitude concerning muon
neutrino contribution. For the type I and II 2HDMs, contributions are above
of the standard model one. It owes to our anzats for mass behavior where tau
neutrino is the heaviest, and also because of the tau lepton mass. For the
type II model, the new physics contribution is only two orders of magnitude
lower on the experimental threshold, for our better case $\tan\beta =68$ and
a mass of charged Higgs of 500 GeV with a contribution close to $8\times10^{-12}$ $\mu_{B}$. 
The type III-scenario reproduces a best
contribution close to $1\times10^{-14}$ $\mu_{B}$ ($\lambda\to1$ and $%
m_{H^{\pm}}=300$ GeV). This case will be one the first parameter spaces to
be constrained with the following experimental values for electromagnetic
form factors.

\item Extracting relevant terms in Eq. (\ref{2HDM_V_FM}), it is possible to
estimate the discrepancies between SM and new physics coming from 2HDM
through the following ratio for MDM contributions

\begin{equation}
R_{\nu }=\frac{\Delta \Lambda _{2HDM}^{\nu _{i}}}{\Lambda _{SM}^{\nu _{i}}}%
\sim O(1)\frac{\frac{m_{l_{k}}m_{l_{i}}^{2}}{m_{H^{\pm }}^{2}}\Xi ^{2}}{%
m_{\nu _{i}}}.  \label{R}
\end{equation}%
Here $\Xi $ is related with the couplings attaching to lepton mass $%
m_{l_{i}} $ in each model, how is depicted in Tab. \ref{tab:couplingsPLPR}.
For instance, in the type II (or Lepton Specific) scenario the relative
coupling is $\Xi _{II}=\tan \beta U_{i,k}$. Taking $\tan \beta =10$, $%
m_{H^{\pm }}=300$ GeV (with diagonal elements of $U_{PMNS}$ matrix and for
the neutrino masses shown in Tab. \ref{tab:EMNC} ), we found for the type
II-2HDM

\begin{equation*}
R_{\nu }^{II}\sim \left\{ 
\begin{array}{c}
O\left( 10^{-3}\right) \text{ with }\nu _{i}=\nu _{e}, \\ 
O\left( 10^{3}\right) \text{ with }\nu _{i}=\nu _{\mu }, \\ 
O\left( 10^{7}\right) \text{ with }\nu _{i}=\nu _{\tau }.%
\end{array}%
\right.
\end{equation*}

\item For the type I (or Flipped) case, it is just necessary to make the
change $\tan \beta \rightarrow \cot \beta $ in the $\Xi $ couplings. For the
same values in this parameter space, we get values for $R_{\nu }^{I}$
roughly belonging to four orders magnitude below to the corresponding
contribution for the type II scenario.

\item For the type III, the ratio for new physics effects is

\begin{align}
R_{\nu}^{III}\sim\frac{\dfrac{m_{l_{i}}}{m_{H^{\pm}}^{2}}\Xi^{2}}{%
G_{F}m_{\nu_{j}}}.
\end{align}

By taking Sher-Cheng anzats (included in the model dependent $\Xi$
coupling), and for values of $\lambda_{l}=0.1$ and $m_{H^{\pm }}=300$ GeV,
the estimates give

\begin{equation*}
R_{\nu }^{III}\sim \left\{ 
\begin{array}{c}
O\left( 10^{-4}\right) \text{ with }\nu _{i}=\nu _{e}, \\ 
O\left( 10^{1}\right) \text{ with }\nu _{i}=\nu _{\mu }, \\ 
O\left( 10^{4}\right) \text{ with }\nu _{i}=\nu _{\tau }.%
\end{array}%
\right.
\end{equation*}

It allows seeing as the structure used by anzats suppresses values of MDMs.
Besides, the form of $R^{III}_{\nu}$ show us as higher values of parameters $%
\lambda^{\prime }s$ shall be the first couplings in to be constrained in
future experiments. This fact will lead to new bounds to FCNC couplings in
the lepton sector.
\end{itemize}

\subsection{Neutrino specific scenario}

In the neutrino specific scenario, the contribution to the MDM in each case
of flavor neutrino is strongly sensitive to the VEV for the second doublet
(which is the one giving mass to the neutrinos). The value of MDM is due to
the structure of the coupling, that depends directly on the neutrino mass,
weakened even more by the neutrino mass hierarchy used. These values are
still far from the current experimental limits. Nevertheless, they are above
of the SM contribution. Besides, in the case of $\nu _{e}$ and $\nu _{\mu }$%
, MDMs of type I and II 2HDMs are below of neutrino specific contributions.
This effect is because the Yukawa structure of the neutrino specific model,
which has the $m_{\nu }/v_{2}$ ratio; meanwhile for other models couplings
between neutrinos and charged Higgs depend on the neutrino masses uniquely.

\begin{figure}[tph]
\centering
\textbf{{\small Neutrino specific \hspace{6.0cm} Neutrino specific}} %
\includegraphics[scale=0.325]{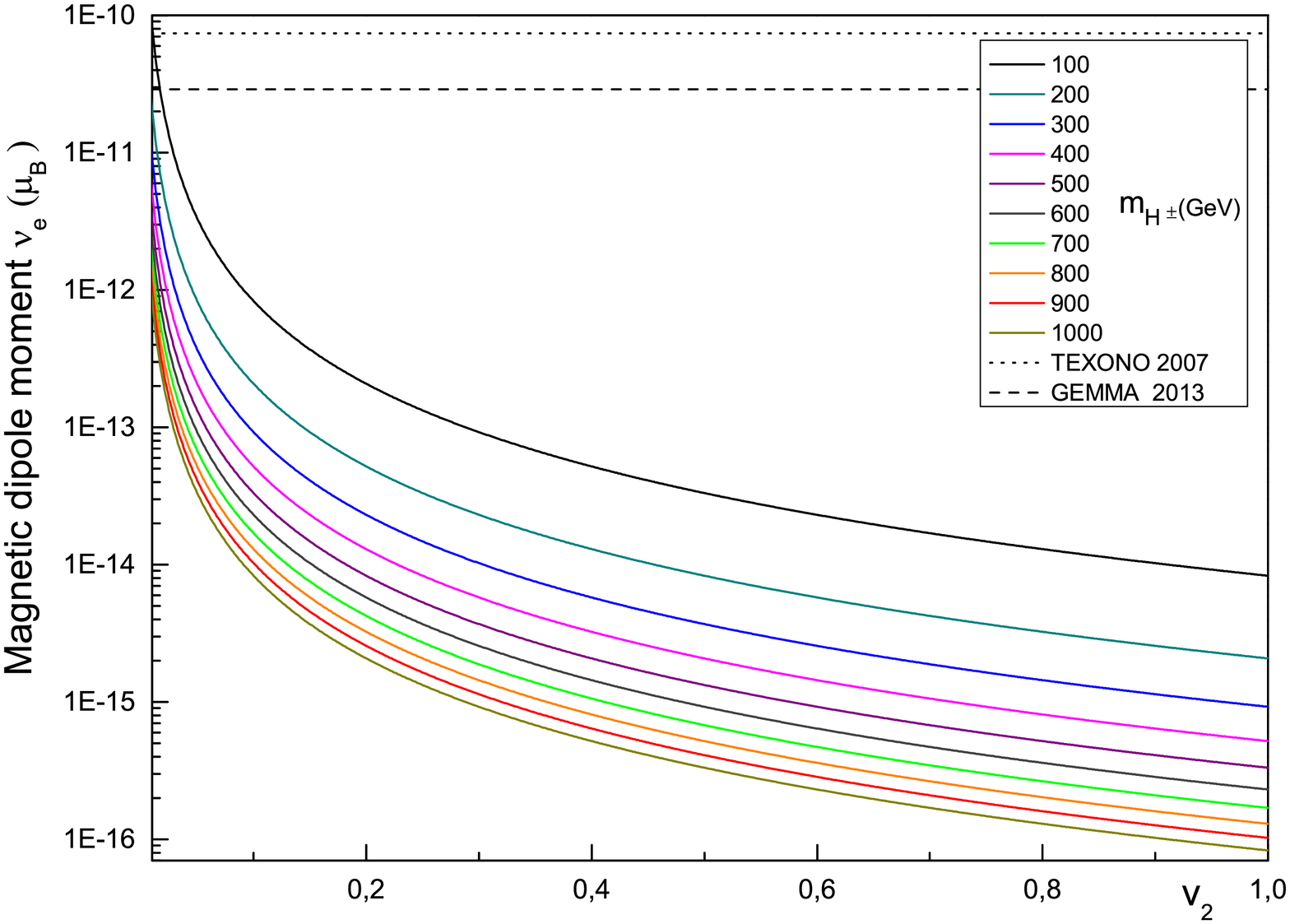}%
\includegraphics[scale=0.325]{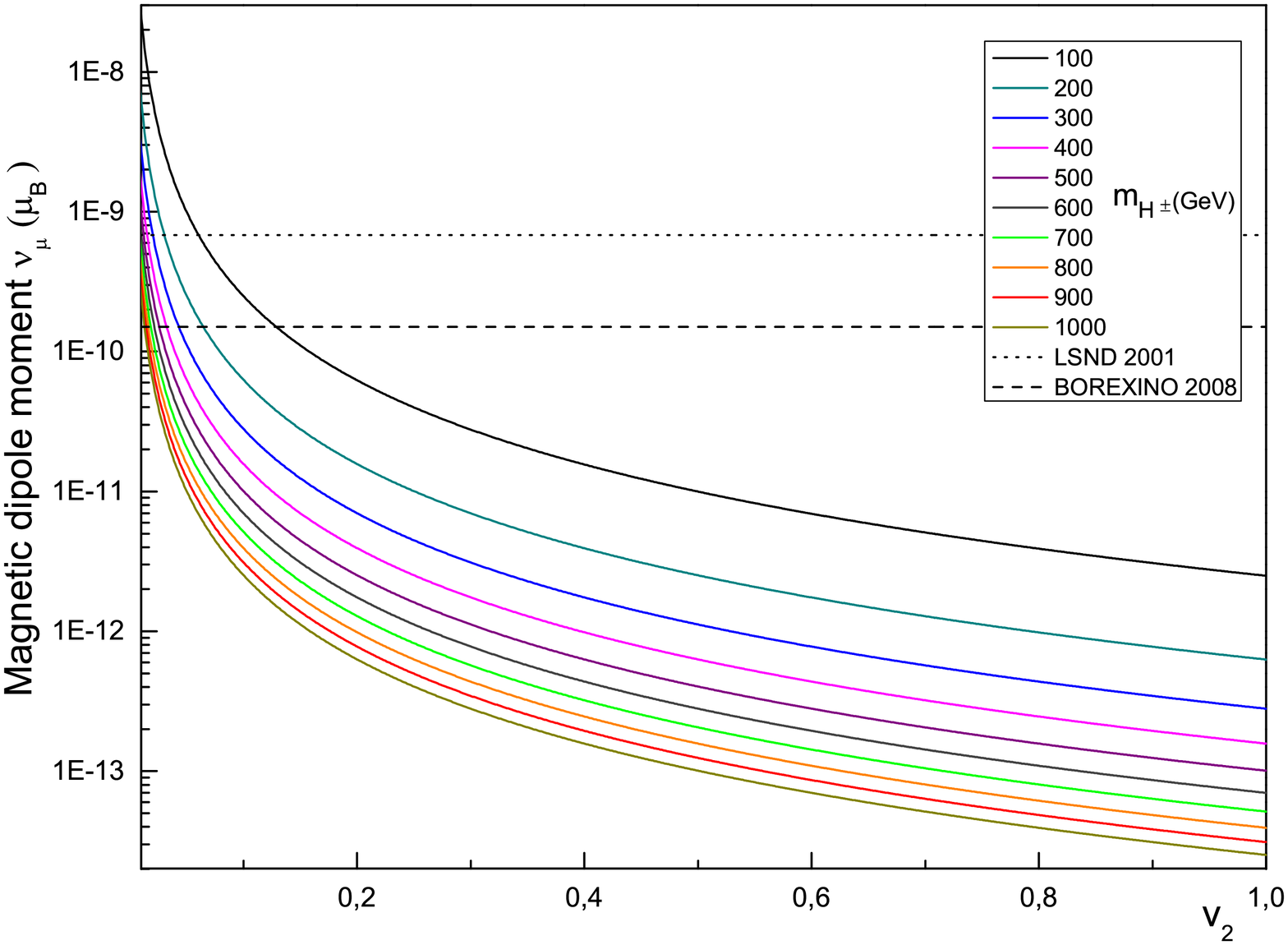} \vspace{-0.4cm}
\caption{\textit{Contribution to the magnetic dipolar moment for (\textbf{%
Left}) electron and (\textbf{Right}) muon neutrinos coming from the neutrino
specific-2HDM with masses of charged Higgs sweeping between }$(100-1000)$%
\textit{\ GeV. The horizontal dotted lines make reference to the
experimental thresholds for muon neutrino at 90$\%$ C.L. Blue line
identifies the lower bound for vacuum expectation value }$v_{2}$\textit{\
based on Yukawa couplings perturbativity.}}
\label{specificI}
\end{figure}

In the case of electron neutrino (see Fig. \ref{specificI}-\textbf{\textit{%
Left}}), exists a significant contribution to the VEV of the second doublet
less to $0.2$ eV, which however is far from the experimental limits of about
two orders of magnitude. Nevertheless, remaining contributions consistent
with experimental thresholds are also above of the SM one by seven-nine
orders of magnitude.

From Fig. \ref{specificI}-\textbf{\textit{Right}}, we see as the
contribution for $\nu_{\mu}$-MDM overpasses the SM-MDM in several order of
magnitude. Experimental thresholds limit the MDM-value when $v_{2}<0.18$ eV
and $m_{H^{\pm}}=100$ GeV. Higher values of charged Higgs mass are allowed
when $v_{2}$ is even lower. For instance, for $m_{H^{\pm}}=300$ GeV, VEV
suppressed satisfy $v_{2}<0.075$ eV. 
\begin{figure}[tph]
\centering
\textbf{\small Neutrino specific}
\par
\includegraphics[scale=0.325]{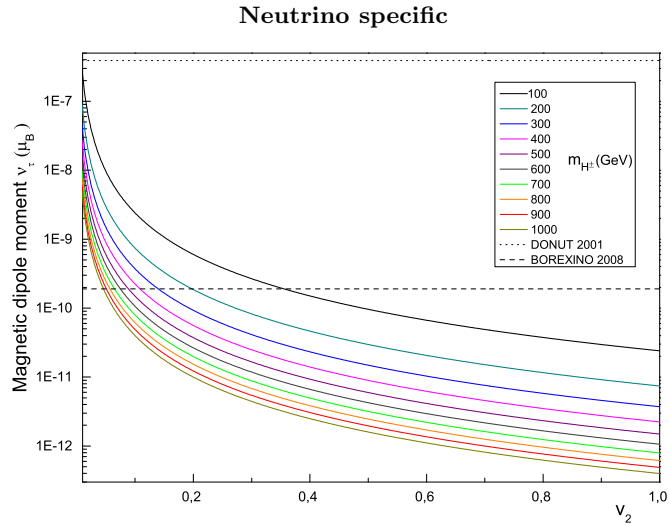} \vspace{-0.4cm}
\caption{\textit{Contribution to the magnetic dipolar moment for tau
neutrinos coming from the neutrino specific-2HDM with masses of charged
Higgs sweeping between }$(100-1000)$\textit{\ GeV. Unitarity constraints
from scattering processes put a limit value for charged Higgs close to }$700$%
\textit{\ GeV \protect\cite{Sher}. The horizontal dotted lines make
reference to the experimental thresholds for tau neutrino at 90$\%$ C.L.
Blue line identifies the lower bound for vacuum expectation value }$v_{2}$%
\textit{\ based on Yukawa couplings perturbativity.}}
\label{specificII}
\end{figure}

For tau neutrinos (see Fig. \ref{specificII}), we have one of the most
promissory scenarios, due to the proximity of the experimental thresholds.
These cases are only one order of magnitude below of measurements limits,
for $v_{2}\leq1$ eV. Hence, Neutrino specific could be a plausible scenario
where new physics could be constrained when the bound of MDM becomes
enhanced employing new precision experiments for neutrino phenomenology.

To describe the discrepancies between new physics from neutrino specific
model and SM contributions quantitatively, and quoting relation (\ref{R})
for other 2HDM types, we define the ratio

\begin{align}
R_{\nu }=\frac{\Delta \Lambda _{2HDM}^{\nu _{i}}}{\Lambda _{SM}^{\nu _{i}}}%
\sim \frac{\frac{m_{l_{k}}m_{v_{i}}^{2}}{m_{H^{\pm }}^{2}v^{2}}\Xi ^{2}}{%
G_{F} m_{\nu _{j}}}.
\end{align}

For $m_{H^{\pm}}=300$ GeV and $v_{2}=0.1$ eV, we find for each type of
neutrino

\begin{equation*}
R_{\nu }^{ns}\sim \left\{ 
\begin{array}{c}
O\left( 10^{6}\right) \text{ with }\nu _{i}=\nu _{e}, \\ 
O\left( 10^{8}\right) \text{ with }\nu _{i}=\nu _{\mu }, \\ 
O\left( 10^{10}\right) \text{ with }\nu _{i}=\nu _{\tau }.%
\end{array}%
\right.
\end{equation*}

However in these particular cases, as we approach near to the experimental
threshold, Yukawa couplings are also reaching the perturbativity limit.
Indeed, these limits are established to avoid divergences in the
renormalization group equations. The reason is that the value of Yukawa
coupling (square) cannot exceed the bound of $8\pi $ since beyond this limit
the chance to find out some Landau pole in energy couplings evolution is
significantly greater \cite{Kanemura}. This value translates into a lower
bound of $v_{2}$ for the respective neutrino mass present in the normal
hierarchy. In the case of $\nu _{\tau },$ for instance, $v_{2\text{min}%
}=0.015$ eV. This VEV is the starting point of all scannings. Thus lower
values of $v_{2}$ overpassing the threshold corrections are all
non-perturbative contributions to MDM. Furthermore, in particular cases,
contributions to MDMs can be constrained by threshold experimental limits,
as for instance from $\nu_{\tau}$ contributions, the fact of $v_{2}>0.4$ eV
for values are allowed at 90 $\%$ C.L. for all charged Higgs masses. By
allowing smaller numbers of $v_{2}$ without to reach the perturbative limit,
the constraints also exclude several values of charged Higgs masses.

From the latter points, we can see as the neutrino specific scenario is
converted into a promising scenario to explain the MDMs and its intimate
relation with neutrino mass.

\section{Concluding remarks\label{sec:conclusions}}

The neutrino magnetic moment provides a tool for exploration of possible
physics beyond the Standard Model. Although the value of the magnetic moment
is suppressed by the smallness of the neutrinos masses, the contribution
from new physics parameters (such as masses and mixing angles) could become
relatively significant. In particular, for the 2HDMs, we evaluated the
contributions coming from the insertion of the charged Higgs bosons into the
loops. Our results show that for the type I and type II 2HDMs (scenarios
without FCNC's), the total contribution is located from the threshold of
experimental detection in the case of electron neutrinos, obtaining a
maximum contribution about three orders of magnitude below of the SM values
(type II-2HDM). In the scenario of muon neutrinos, the total contribution
produces comparable values with the SM contributions for the parameters of
the model type I, with the higher contributions coming from the case of
model type II. Finally, such values are much higher for tau neutrinos, but
those contributions are much stronger for the model type II with $\tan \beta
>>1$ and $m_{H^{\pm }}=500$ GeV. The last scenario for $\nu_{\tau }-$MDM is
only two order of magnitude below of the threshold experimental given by
BOREXINO experiment.

Despite the type-III 2HDM shows a new scenario to interpret MDM as a result
of FCNCs, its contributions are tiny, even compared with the experimental
threshold values. This effect is due to the suppression yielded by the
Cheng-Sher Anzats which defined that FCNCs couplings satisfy the same
hierarchical structure for leptonic mass matrices. This fact makes that
FCNCs for neutrinos and charged leptons become suppressed by the same
EW-scale.

Although the models without FCNCs converts into plausible scenarios for
electromagnetic form factors, such frameworks are based on unnatural terms
for neutrino masses. To circumvent this issue, we consider a simple 2HDM
that incorporates viable masses for neutrinos. It is done by choosing a VEV
in the eV scale (or even less) for the second doublet ($v_{2}$ ), which is
coupled only to Dirac neutrinos. From this assumption, plausible masses
could be explained if other parameters in the model, e.g. $m_{12}^{2}$, take
appropriate values. Besides this neutrino-specific model has other
phenomenological motivations (e.g. $\mu \rightarrow e\gamma $ decays,
cosmology and neutrino phenomenology), in our case, this has been used to
determine the nature of MDM and the relation with small sizable masses in
the neutrino sector. For instance, neutrino specific-2HDM gives
characteristic scales up to $10^{-7}$ $\mu _{B}$ for $\nu_{\tau}$-MDM (even
respecting the perturbative limit on Yukawa couplings). Higher values in MDM
for the neutrino-specific model are due to Yukawa couplings structure, which
scales with the ratio between neutrino mass and $v_{2}$. The last fact is a
radical difference with the remaining Yukawa couplings structures of other
considered 2HDMs. Under the assumption of normal ordering for neutrinos,
these scales for MDMs are above of SM contributions and reach the threshold
measurement, making of neutrino specific a great scenario to constrain new
physics in future experiments.

An important aspect is the close relationship of MDMs with the neutrino mass
and with the respectively charged lepton mass. In the type I-III 2HDMs, this
can be clearly appreciated when the contribution due to the tau neutrino is
compared concerning the muon and electron neutrinos ones, further
strengthened by the normal hierarchy that we are assuming. Another relevant
fact is that if we found magnetic dipole moment for neutrinos, such a
measure would be highly sensitive to the type of 2HDM. However, at precision
order we are working to neutrino MDMs, Lepton-specific analyses are the same
that type II ones. Moreover, the flipped model shares the same remarks that
case for type I-2HDM. Therefore, we have studied the influence on
neutrinos-MDMs of all possible 2HDMs with natural flavor conservation.

In conclusion, $\nu _{\tau }$-MDM could be converted into a plausible
framework to constrain new physics scenarios contributing to electromagnetic
form factors. Particularly, in the light of future experiments for coherent
neutrino-nucleus scattering, which are expected to improve all bounds on
neutrino electromagnetic properties \cite{ValleNew}. Particularly, and
despite perturbativity could be in conflict with lower values of the natural
VEV $v_{2}$, the neutrino specific model would be a significant benchmark to
introduce these effects of new physics, at the same time that neutrino
masses are at one viable scale. This mechanism leads to a well-motivated
study of 2HDMs itself or implemented under more robust theories (e.g. $B-L$
gauge extended plus 2HDMs or SUSY models) containing a compatible
explanation of smallness for neutrino masses and the influence of $CP$
phases in neutrino-phenomenology. In these scenarios, our results are also
relevant in the regimen where new gauge bosons or new fermions in those
theories become decoupled.

\section{Acknowledgments}

We acknowledge financial support from Colciencias and DIB (Universidad
Nacional de Colombia). Carlos G. Tarazona also thanks to Universidad Manuela
Beltran as well as the support from DIB-Project with Q-number \textbf{%
110165843163} (Universidad Nacional de Colombia). A. Castillo, J. Morales,
and R. Diaz are also indebted to the \emph{Programa Nacional Doctoral de
Colciencias} for its academic and financial aid.

\end{document}